\documentclass[aip, jcp, reprint]{revtex4-2}
\usepackage[english]{babel}

\usepackage{graphicx}
\usepackage{subfig}
\usepackage{mathtools}

\usepackage[version=3]{mhchem} 
\usepackage{dcolumn}
\newcolumntype{d}[1]{D{.}{\cdot}{#1} }
\usepackage{bm}
\usepackage{xcolor,soul}
\usepackage{float}
\usepackage{physics}
\usepackage{tikz}
\usepackage{braket}
\usetikzlibrary{decorations.markings}
\usepackage{orcidlink}

\usepackage{overpic} 

\usepackage{mathpazo} 

\usepackage{siunitx}

\usepackage{verbatim}

\begin{document}
\title{
Quantum dynamic simulations of triplet formation in an effective model of Y6 dimers}
\date{\today} 

\author{Isabel Creed\,\orcidlink{0009-0001-3930-7873}}
\affiliation{Department of Chemistry, Imperial College London, Exhibition Road, London  SW7 2AZ, UK}

\author{Lucy J. F. Hart}
\affiliation{Department of Physics, Imperial College London, Exhibition Road, London  SW7 2AZ, UK}

\author{Pranay Venkatesh}
\affiliation{Department of Chemistry, Imperial College London, Exhibition Road, London  SW7 2AZ, UK}
\affiliation{Department of Chemistry, University of Colorado, Boulder, Boulder, CO, 80309, USA}

\author{Tom Ward}
\affiliation{Department of Chemistry, Imperial College London, Exhibition Road, London  SW7 2AZ, UK}

\author{Jarvist Moore Frost\,\orcidlink{0000-0003-1938-4430}}
\email[Electronic mail:]{jarvist.frost@imperial.ac.uk}
\affiliation{Department of Chemistry, Imperial College London, Exhibition Road, London  SW7 2AZ, UK}
\affiliation{Department of Physics, Imperial College London, Exhibition Road, London  SW7 2AZ, UK}


\begin{abstract}
We construct a five-state model for photoexcitation in Y6 (BTP-4F) dimers, and
then solve the non-adiabtic dynamics using the Hierarchical
Equations of Motion (HEOM) method. 
We find that triplets are populated mainly via a transiently excited
\textit{intermolecular} charge-transfer singlet to triplet Frenkel exciton
route; this route is not available to the monomer. 
Analysis of one-particle transition density matrices suggests that the
charge-transfer states are spatially distinct to the Frenkel exciton states, indicating
that the large spin-orbit-coupling for this transition is due to it being
permitted by an associated change in orbital character. 

Aggregation in Y6 therefore directly enables fast and high-yield intersystem
crossing.  
We selenise our model dimers, significantly enhancing spin-orbit-coupling,
which then accelerates this charge-transfer mediated route. 
Looking forwards to simulations on larger aggregates, we show that, though
Marcus theory gives qualitatively correct dynamics, the long-time yields are
incorrect due to it missing quantum recurrences. 
Instead, we show that the recently developed memory-kernel
projector\cite{Gestsson2025-ez} method can produce semi-classical rates
directly from the HEOM equations which lead to quantitatively correct
dynamics and yields. 

\end{abstract}

\maketitle

\section{Introduction}\label{introduction} 

Predicting rates of internal-conversion (IC) and intersystem crossing (ISC) are of foundational interest in theoretical photochemistry. 
Modelling these processes is essential to interpret transient photo measurements of organic electronic materials. 

One system of current technical interest is Y6 (BTP-4F)\cite{Yang2021}. 
This flat A-DA'D-A push-pull structure shows some unusual photophysics. 
There is some evidence that free charges may be directly generated upon photoexcitation
without an applied bias\cite{price2022free}; certainly the energetic cost of exciton
dissociation into free charges is small\cite{Zhu2022,Hart2026}. 
The origin and role of triplet excitations in organic photovoltaics remain debated. 
 
Izawa and Hiramoto\cite{izawa2021efficient} showed that Y6, in a bilayer with evaporated rubrene, upconverts infrared light into visible (via triplet-triplet annihilation; two photons in for one photon out) with the highest efficiency for a solid-state device. 
A central question is where these triplets come from: Izawa and Hiramoto offered circumstantial evidence that they result from charge-separation and then recombination at the Y6/rubrene interface (producing 3/4 triplets in the usual spin ratios); the obvious alternative hypothesis is that they are directly generated via ISC from the photoexcited singlet of Y6. 

If the triplets are mainly produced by charge-separation and recombination, one anticipates greater upconversion power conversion efficiency, as you do not have to pay the exchange energy penalty of intersystem crossing from $S_1$ to $T_1$. 
If the triplets are mainly produced by intersystem crossing, then there is a simple molecular design rule: substitute heavier atoms, as the spin-orbit-coupling (SOC) increases with the nuclear charge $Z$ as $\mathcal{O}(Z^2)$ leading to the intersystem crossing rate increasing as $\mathcal{O}(Z^4)$.

Motivated by the approach of Greyson et al.\cite{Greyson2010} and Parker et al.\cite{Parker2014} in constructing effective models to explain \textit{singlet fission}, our long-term goal is to build an \textit{effective model} for \textit{triplet-fusion} upconversion in Y6:rubrene films. 

The concept of intermolecular charge transfer (CT) states\cite{si2023photo, price2022free,izawa2021efficient, lan2024correction} are often invoked to explain unusual photophysics of Y6. 
As these charge-transfer states are intermolecular, we need a structural model of packing, and must, at the very least, consider dimers as a model for the solid state. 
Early studies required this to come from molecular dynamic simulations\cite{Zhang2020},
but more recent work\cite{giannini2024role, Ward2026} often use relatively recent high quality crystal structures\cite{Xiao2020} (CSD code OHUBUR). 
Here we use dimers extracted from this crystal structure, following the approach and numbering scheme of Giannini et al.\cite{giannini2024role}. 

Recently\cite{Ward2026}, we discussed the relative ordering of the CT and FE singlet and triplet states in the various Y6 dimers and the impact of FE-CT hybridization on the character of these states, and the impact of this character on triplet formation. 
We now try and directly simulate this process.

We construct an effective Y6 dimer model for the solid-state, and consider the effect of selenisation (to increase SOC) on the predicted photophysics. 
Our Y6Se system is the same dimer structures as Y6 with the two outer sulphurs replaced with selenium (the backbone of T9SBN-F, the synthesis of which was first reported by Jiang et.~al\cite{Jiang2022} in 2022), and no further relaxation. 
Our five-state effective Hamiltonian is parametrised directly from matrix elements calculated with density functional theory on the actual Y6 (and Y6Se) dimer structures. 
Our model is complete with a judicious choice of two phonon baths to best represent the dissipative environment of Y6 (Y6Se) in the solid state. 
We then simulate the dynamics with Hierarchical Equations of Motion (HEOM), and compare to rates obtained from Marcus theory and from projecting out effective rates from the HEOM memory kernel. 

We find that Y6 dimers directly produce triplets on a nanosecond timescale, via ISC from the singlet Charge Transfer (CT) to triplet Frenkel Exciton (FE) states. 
This route is only possible in the solid-state where intermolecular CT states are present. 
Selenisation of Y6 causes an order of magnitude increase in the rate of triplet formation as expected for triplets formed by intersystem crossing, with the triplets still formed by the same mechanism as for the Y6 dimers. 

Souza et al.\cite{Souza2025} recently proposed that in the Y6 monomer, extreme distortion of the molecule around the key tortion\cite{Kupgan2021} in the $S_1$ excited state (to $90^o$) leads to large spin-orbit coupling, which then explains the measured ISC rate. 
We do not believe that size of distortion is possible in the solid-state, where it will be severely sterically hindered by the close $\pi-\pi$ stacking. 
Their analysis in the Frank-Condon region (with and without Herzberg-Teller corrections) produces a rate insufficient to explain the data. 

\section{Theory and Methods}

HEOM is computationally demanding\cite{Tanimura1989}, as it directly describes the system-bath interactions via a hierarchy of coupled differential equations. 
This in practice means that the computational time and memory scales as a factorial\cite{Ke2023} of the number of environmental modes, size of your system and with convergence parameters $L$ and $K$ (see Section \ref{subsec:HEOM}).
Studying a five-state effective model enables us to stretch to using a sophisticated and realistic phonon bath. 

In the future we want to be able to bridge the gap to scalable models (i.e. to simulate exciton diffusion, whole device models), so we are motivated to construct Master equations of classical rates. 
One approach is to directly calculate Marcus\cite{Marcus1965} rates from our Hamiltonian. 
A more sophisticated and recently developed approach is to use Nakajima-Zwanzig projection operator formalism to project out time-independent effective rates for a generalised Master equation from the HEOM memory kernel\cite{Gestsson2025-ez}.  

In order to treat the nonadiabatic dynamics accurately, we first simplify our model.

\subsection{Dimer effective Hamiltonian}

We take our six dimers (and numbering scheme 1--6) from the recent work by Giannini et al.\cite{giannini2024role}. 
These `contact pair' dimers were extracted from a solved crystal structure\cite{Xiao2020}, with a distance cutoff of \SI{3.5}{\AA}. 
Transfer integrals become significant and start to affect the excited state energies at these small separations\cite{olaya2011energy}. 

We then project onto a five-state \textit{effective} model, with the general Hamiltonian shown in Figure \ref{ham}. 

\begin{figure*}[t]
\begin{equation} \hat H_{\text{eff}} = 
\begin{bmatrix*}[l]
    E_{S_0} & V_{S_0 \leftrightarrow  S_{FE}} & V_{S_0 \leftrightarrow S_{CT}} & V_{S_0 \leftrightarrow   T_{FE}} & V_{S_0 \leftrightarrow  T_{CT}} \\
    V_{S_0 \leftrightarrow  S_{FE}} & E_{S_{FE}} &    V_{S_{FE} \leftrightarrow  S_{CT}}  & V_{S_{FE} \leftrightarrow   T_{FE}} & V_{S_{FE} \leftrightarrow  T_{CT}}  \\
    
     V_{S_0 \leftrightarrow  S_{CT}}  & V_{S_{FE} \leftrightarrow  S_{CT}} &E_{S_{CT}} & V_{S_{CT} \leftrightarrow  T_{FE}}  & V_{S_{CT} \leftrightarrow T_{CT}}  \\
     
    V_{S_0 \leftrightarrow   T_{FE}}  & V_{S_{FE} \leftrightarrow  T_{FE}}  & V_{S_{CT} \leftrightarrow  T_{FE}}  &  E_{T_{FE}}&  V_{T_{FE} \leftrightarrow  T_{CT}} \\
    V_{S_0 \leftrightarrow  T_{CT}} & V_{S_{FE} \leftrightarrow  T_{CT}} &  V_{S_{CT} \leftrightarrow  T_{CT}}  &  V_{T_{FE} \leftrightarrow  T_{CT}}  &  E_{T_{CT}}   \\
\end{bmatrix*}
\label{ham}
\end{equation}
\caption{General five-state Hamiltonian matrix used in our dimer modelling.}
\end{figure*}

Our underlying electronic structure method is B3LYP global-hybrid density functional theory (DFT), with a modest 6-31G(d, p) basis set. 
We recently found that this combination fortuitously describes the key excited states in these dimers accurately and with correct ordering\cite{Ward2026}, in reference to higher level and optimally tuned range-separated hybrid calculations. 
This combination of functional and moderate basis set is known to be well behaved and with reliable convergence: we considered this feature extremely important as we are combining matrix elements calculated with very different approximations, and so the precision achieved in our calculations is more important than the overall accuracy of the method. 
We use the Tamn-Dancoff Approximation (TDA) consistently in our time-dependent DFT (TDDFT) calculations, as we found\cite{Ward2026} this led to well behaved ordering of the excited states. 
Initially we had a concern that in our selenium calculations we might need to consider relativistic contributions to the electronic structure. 
After some testing (see Supplementary Information) we found that the effect was minor, and just quantitative in terms of excited state energy - mainly from need to turn the TDA approximation off in the relavistic calculations - rather than significantly perturbing the wavefunction. 

For both Y6 and Y6Se, our Hamiltonian (Figure \ref{ham}) contains five effective states: the singlet ground state ($S_0$), the singlet and triplet effective Frenkel Exciton ($S_{FE}$,$T_{FE}$) and the Charge Transfer ($S_{CT}$,$T_{CT}$) states. 
To define these effective states, we first characterise the excited states by calculating a charge-transfer metric from the single-particle transition density matrix projected onto the respective molecular fragments (see S.\ref{theodore} in the Supplementary Information). 

The energies of these effective states are given by the arithmetic mean of the dimer excitation energies. 
Pooling between off-diagonal matrix elements is done by taking the average in \textit{quadrature}, as this then respects Fermi's golden rule (FGR) in correctly combining semi-classical rates. (See Supplementary Information for the specific summations.)

For the FE--CT couplings between states of the same spin-multiplicity, we make a frozen orbital approximation and construct the pathways via the four different single-electron transfers. 
If we label the two monomer units as A and B, the four pathways are $A^+B^- \rightarrow A^*B$, $A^-B^+ \rightarrow AB^*$, $A^-B^+ \rightarrow A^*B$ and $A^+B^- \rightarrow AB^*$. 
The first two processes are electron transfer ($V_e$, approximated as the effective LUMO-LUMO coupling), and the latter two processes are hole transfer ($V_h$, approximated as the effective HOMO-HOMO coupling). 
We estimate $V_h$ and $V_e$ with a counterpoise-corrected projection method\cite{Kirkpatrick2007, Baumeier2010}, which projects the monomer HOMO (LUMO) orbital through the orbitals of the dimer. 
The underlying DFT (B3LYP/6-31g* on the individual monomers, and dimer) are undertaken in Gaussian 16 \cite{g16}, due to compatibility with our group codes.

For the spin-orbit-coupling between singlet and triplet channels, we directly take the quadrature sum of the spin-orbit-coupling matrix elements (SOCME) between the different excited states, mapped to the relevant effective FE and CT states. 
This includes the SOCME between $T_1$ and $S_0$ which drives phosphorescence. 
The calculation in Orca uses the SHARK Integral Package\cite{Neese2020}, which uniquely provides analytic integrals for the $L \cdot S$ operator, making convergence reliable even with a modest basis set and standard DFT integration grids. 

We do not calculate explicitly matrix elements between the singlet states and the ground states in our model Hamiltonian. 

\subsection{Marcus Theory}\label{ssec:Marcus}

The celebrated Marcus rate theory \cite{Marcus1965} approach has been applied to modelling organic electronic materials for many decades. 
Transfer processes are modelled as being rare and driven by thermal fluctuations, and occur incoherently. 
These rates correctly include nuclear quantum effects (in a high-temperature limit, therefore the frequency of the Bath modes are considered small; the Bath is structureless). 
The rates are perturbative, so assume that the couplings are small relative to the reorganisation energy. 

Key for our work is that these rates are also Markovian - the Bath thermalises quickly relative to the rate of hopping, so the system retains no memory of the previous motion. Within this framework, the Marcus rates, $\Gamma_{nm}$ are given by

\begin{equation}
    \Gamma_{nm} = \frac{2\pi}{\hbar}\frac{|H_{nm}|^2}{\sqrt{4 \pi \lambda k_B T}} \exp(-\frac{(\lambda + \Delta G)^2}{4\lambda k_B T}) , 
\end{equation}
with $H_{nm}$ and $\Delta G= H_{nn}-H_{mm}$ from our effective system Hamiltonian $\hat H_{\text{eff}}$, and $\lambda$ the reorganisation energy (system-bath coupling). 
As is standard in this area of modelling, we use an empirical (and apparently larger than what is demonstrated in our electronic structure calculations) reorganisation energy of \SI{0.5}{\electronvolt}. 
Most likely, this empirically corrects for the fact that we are taking Marcus theory out of the perturbative regime. 

From these rates we construct a Master equation rate matrix as 
\begin{equation}
\frac{dP_n}{dt} = \sum_{m \neq n} \left[\Gamma_{mn} P_m(t) - \Gamma_{nm} P_n(t)\right] , 
\end{equation} 
where the $\Gamma_{mn}$ are as above. 
This Master equation can then be trivially numerically propagated with standard ordinary differential equation solvers. 

\subsection{Hierarchical equations of motion (HEOM)}\label{subsec:HEOM}

Established semi-classical rate theories rely on perturbative treatments and, being Markovian, cannot describe quantum recurrence or quantum coherence effects (particularly via transiently occupied states). 

We therefore apply the formally exact HEOM method\cite{Tanimura1989}. 
Rather than integrate out the Bath entirely, HEOM constructs the system-bath entanglement (history) via a hierarchy of auxiliary density operators. 
HEOM propagates the exact non-Markovian dynamics up to a truncation depth $L$, beyond which bath memory effects are assumed to instantaneously decay. We achieved convergence at a hierarchy depth of $L=3$ (convergence tests are described in the Supplementary Information, see Figure \ref{HEOM_Con}). 

With our simplified five-state system Hamiltonian ($\hat H_{\text{eff}}$, see Figure \ref{ham}) we can computationally afford to model the dissipative Bath with some care. 
We therefore simultaneously couple to two, independent, phonon baths. 

Our first phonon bath is Drude-Lorentz (Debye), which describes the slow (typically intermolecular) modes, which are assumed to be highly anharmonic and frictional. 
We take the a reorganisation energy of $\lambda=0.034$ eV\cite{giannini2024role} and a relaxation rate $\gamma=0.050$ eV ($\simeq 2 k_B T$).

Our second phonon bath is an \textit{under-damped} Brownian oscillator (Lorentzian). 
This represents the high-frequency (typically intramolecular) modes, which can provide multi-phonon resonances and energy exchange. 
For this mode we take $\lambda_u=0.052$ eV\cite{giannini2024role}; and somewhat empirically choose our damping rate of $\gamma_u=0.015$ eV with an effective peak frequency of $u_{max}=0.160$ eV to get convergent dynamics.

The Matsubara expansion of the thermal bath correlation function, is truncated at $K=3$ for the slow Drude-Lorentz bath and $K=1$ for the rapidly decaying under-damped bath. 

Finally, we incorporate phenomenological Lindblad dissipaters enabling (fluorescent) relaxation from the singlet excited states to the ground state ($S_0$). 
Based on experimental transient absorption data\cite{Zou2020}, we apply a decay rate of $10^9 \text{ s}^{-1}$ for the $FE(S) \to S_0$ transition and $10^8 \text{ s}^{-1}$ for the $CT(S) \to S_0$ transition. 

For consistency, all HEOM simulations presented here were undertaken in the efficient and reliable Julia package \textsc{HierarchicalEOM.jl}\cite{Huang2023}, but during development we also made key use of \textsc{pyrho} \cite{pyrho}, and \textsc{QuantumDynamics.jl}\cite{Bose2023}.

\subsection{Effective rates from HEOM}\label{ssec:HEOMRates}

The Nakajima-Zwanzig projection operator constructs a generalised Master rate equation by projecting the full density matrix onto the population of a relevant space. 
Gestsson et al.\cite{Gestsson2025-ez} recently showed how this can be used explicitly to calculate an effective time-independent rate constant from the time-dependent and non-Markovian full HEOM dynamics. 

The rate between the states $\ket{I}$ and $\ket{J}$ is given by
\begin{equation}
    \Gamma_{I\rightarrow J}= \braket{J|\mathcal{PLGQLP}|I} \braket{I|J}
\end{equation}
where $\mathcal{P}$ is the projection operator (defined below), the irrelevant space which we are projecting out is $\mathcal{Q} = \mathcal{I} - \mathcal{P}$ (where $\mathcal{I}$ the identity operator), and the resolvent (propagator Green's function of $\mathcal{Q}$ space) $\mathcal{G}$ is 
\begin{equation}
\mathcal{G} = -\lim_{\varepsilon\to0^+}\frac{1}{\mathcal{QL}-\varepsilon\mathcal{I}}
\end{equation}
Here the limit $\varepsilon \to 0^+$ can be understand as a Laplace transform evaluated at $s=0$, which integrates out the time dependence to give a steady-state result. 
Therefore, this transform integrates out any coherent beating dynamics, but does include quantum recurrences, and quantum coherence of transiently occupied states. 
We construct the projection operator to project onto the effective states of $H_\mathrm{eff}$ (see Supplementary Information). 

We did not apply the Ishizaki-Tanimura correction term (for the incoherent Lindblad dissipaters), but rather replaced these dissipaters with empirical fluorescent terms of $\gamma_{S_1}=1\times10^9$ and $\gamma_{S_{CT}}=1\times10^8$. 
In order to get accurate dynamics in the long time limit, we increased $K$ from 3 to 7 for the Drude-Lorentz bath.

\section{Results and discussion}

\subsection{Effective model}

\begin{table}[t]
\begin{ruledtabular}
\begin{tabular}{ l c c c c c c } 
\textbf{Y6 Dimers} & \textbf{D1} & \textbf{D2} & \textbf{D3} & \textbf{D4} & \textbf{D5} & \textbf{D6} \\ 
\colrule
\multicolumn{7}{l}{\textit{Effective Site Energies (eV)}} \\
$E_{FE(S)}$ & 1.947 & 1.907 & 1.868 & 1.932 & 1.976 & 1.914 \\
$E_{CT(S)}$ & 1.675 & 1.710 & 1.656 & 1.743 & 1.815 & 1.779 \\
$E_{FE(T)}$ & 1.481 & 1.478 & 1.466 & 1.464 & 1.507 & 1.469 \\
$E_{CT(T)}$ & 1.737 & 1.749 & 1.671 & 1.804 & 1.850 & 1.799 \\
\colrule
\multicolumn{7}{l}{\textit{Spin-Orbit Coupling Matrix Elements (cm$^{-1}$)}} \\
CT(1)-CT(3) & 0.131 & 0.081 & 0.046 & 0.143 & 0.052 & 0.046 \\ 
CT(1)-FE(3) & 0.363 & 0.183 & 0.121 & 0.193 & 0.258 & 0.191 \\
FE(1)-CT(3) & 0.330 & 0.136 & 0.175 & 0.156 & 0.081 & 0.152 \\
FE(1)-FE(3) & 0.090 & 0.046 & 0.010 & 0.072 & 0.073 & 0.142 \\
GS-FE(3)    & 1.815 & 2.106 & 1.712 & 1.972 & 2.156 & 1.641 \\
GS-CT(3)    & 0.802 & 0.000 & 0.783 & 0.301 & 0.073 & 0.197 \\
\colrule
\multicolumn{7}{l}{\textit{CT--FE Coupling (eV)}} \\
CT(S/T)-FE(S/T) & 0.108 & 0.089 & 0.059 & 0.047 & 0.070 & 0.050 \\
\end{tabular}
\end{ruledtabular}
\caption{Effective Hamiltonian matrix elements for the six Y6 contact pair dimers. Site energies (eV), spin-orbit coupling matrix elements (SOCMEs) between singlet and triplet states (cm$^{-1}$); and the coupling between charge-transfer (CT) and Frenkel exciton (FE) states of the same multiplicity.}
\label{tab:y6_merged_params}
\end{table}

\begin{table}[t]
\begin{ruledtabular}
\begin{tabular}{ l c c c c c c } 
\textbf{Y6Se Dimers} & \textbf{D1} & \textbf{D2} & \textbf{D3} & \textbf{D4} & \textbf{D5} & \textbf{D6} \\ 
\colrule
\multicolumn{7}{l}{\textit{Effective Site Energies (eV)}} \\
$E_{FE(S)}$ & 1.905 & 1.882 & 1.840 & 1.915 & 1.948 & 1.884 \\
$E_{CT(S)}$ & 1.652 & 1.678 & 1.621 & 1.683 & 1.772 & 1.739 \\
$E_{FE(T)}$ & 1.475 & 1.485 & 1.462 & 1.452 & 1.510 & 1.458 \\ 
$E_{CT(T)}$ & 1.708 & 1.716 & 1.628 & 1.784 & 1.820 & 1.767 \\
\colrule
\multicolumn{7}{l}{\textit{Spin-Orbit Coupling Matrix Elements (cm$^{-1}$)}} \\
CT(1)-CT(3) & 0.862 & 0.427 & 0.252 & 0.979 & 0.476 & 0.322 \\ 
CT(1)-FE(3) & 3.397 & 0.583 & 0.526 & 1.039 & 2.352 & 1.277 \\
FE(1)-CT(3) & 2.089 & 0.260 & 0.540 & 1.093 & 0.479 & 0.444 \\
FE(1)-FE(3) & 0.659 & 0.233 & 0.024 & 0.529 & 0.163 & 0.282 \\
GS-FE(3)    & 8.564 & 9.051 & 9.717 & 9.735 & 9.853 & 9.764 \\
GS-CT(3)    & 5.980 & 0.000 & 1.618 & 0.795 & 3.090 & 1.781 \\
\colrule
\multicolumn{7}{l}{\textit{CT--FE Coupling (eV)}} \\
CT(S/T)-FE(S/T) & 0.099 & 0.104 & 0.053 & 0.069 & 0.078 & 0.052 \\
\end{tabular}
\end{ruledtabular}
\caption{Effective Hamiltonian matrix elements for the six Y6Se contact pair dimers. Site energies (eV), spin-orbit coupling matrix elements (SOCMEs) between singlet and triplet states (cm$^{-1}$); and the coupling between charge-transfer (CT) and Frenkel exciton (FE) states of the same multiplicity.}
\label{tab:y6se_merged_params}
\end{table}

Our effective model is fully specified in Table \ref{tab:y6_merged_params} for the six Y6 dimers, and Table \ref{tab:y6se_merged_params} for the six Y6Se dimers. 

\begin{figure}
    \centering
    \includegraphics[width=\linewidth]{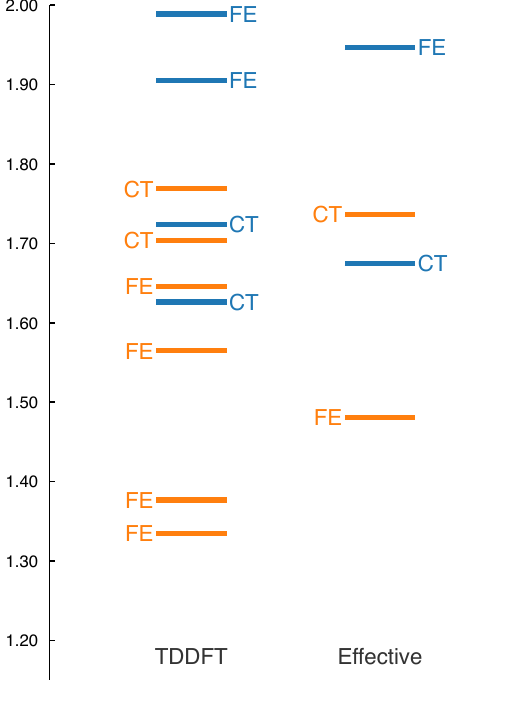}
    \caption{Jablonski diagram (energy units eV) for the D1 dimer. 
    Triplet (orange; FE/CT labels on left) and singlet (blue; FE/CT labels on right) excited states. 
    Underlying TDDFT excitations (left-hand side) are compared to the reduced effective model (right-hand side).
    Effective state energies are the arithmetic mean of the contributing states. 
    }
    \label{Jablonski}
\end{figure}

The Jablonski diagram for the Y6 D1 dimer showing the underlying TDDFT states, and from obtained from diagonalisation of our effective Hamiltonian is shown in Figure \ref{Jablonski}. The correct ordering of the singlet (CT lower than FE) and triplet (FE lower than CT) is reproduced. The inversion of states in this Y6 dimer, also noted in our recent OT-SRSH paper \cite{Ward2026}, is key for the correct dynamics. 

Generally the excited state energies are similar between the different dimers and Y6/Y6Se, and as expected, SOCMEs for Y6Se are an order of magnitude larger than Y6 due to ``the heavy atom effect''. 
The largest SOCMEs are phosphorescence from the triplet excited states to the ground state, but due to the energy gap, this rate is minimal. 

A key observation from these matrix elements is that 
$S_{CT} \leftrightarrow T_{FE}$ and $S_{FE} \leftrightarrow T_{CT}$ SOCMEs are considerably larger than transitions within the same charge-transfer nature. 
We believe this can be understood as a generalisation of El Sayed's rule\cite{Marian2021}: change in orbital location (CT versus FE) drives a change in orbital symmetry, and so CT$\to$FE is more permitted.  

By contrast, the Y6 monomer calculated with the same method exhibits an $S_1$ energy of 1.877 eV and low-lying triplets at 1.311 eV ($T_1$) and 1.560 eV ($T_2$); and SOCMEs between $S_1$ and these triplets which are negligible (0.01 and 0.07 cm$^{-1}$, respectively). 

To understand why the monomer SOCMEs are so small, we follow Mutovska et al.\cite{Mutovska2026} and calculate electron-hole correlation diagrams of the monomer (Figure \ref{monomer_e_h}) and dimers (Figures \ref{D1_e_h_S}-\ref{D4_e_h_T}) singlet and triplet excited states (see Supplementary Information), and see that the monomer singlet and triplet characters are very similar, whereas in the dimers CT and FE states are distinct in spatial extent.  
We recently explained this\cite{Ward2026} as being due to the significantly different delocalization and hybridization of the singlet and triplet states in Y6 dimers. 

To make this quantitative, we define a scalar metric of similarity of the electron-hole plots (states $m$ and $n$) by a simple overlap,  
\begin{equation}
    L_{nm}=\frac{\left(\sum_{I, J} D_{I, J}^m D_{I, J}^n \right)^2}{\left(\sum_{I, J} [D_{I, J}^m]^2\right)\left( \sum_{I, J}[D_{I, J}^n]^2\right)}.  
    \label{Lnmeq}
\end{equation}
The sum is over the electron fragments $I$ and hole fragments $J$.  
This metric is $L_{nm}=1$ if the states are identical,  $L_{nm}=0$ if the states are spatially orthogonal.

Table \ref{Lnmtab} presents the $L_{nm}$ spatial overlap metric for the Y6 monomer and the D1 dimer. 
The monomer's $S_1$ state shares nearly identical electron-hole distributions with its $T_1$ and $T_2$ states ($L_{nm} > 0.91$). 
Transitions between the distinct FE and CT states in the D1 dimer exhibit near-zero spatial overlap ($L_{nm} < 0.01$). 
We propose that these changes in spatial overlap drive a change in orbital character, permitting spin-orbit-coupling between the FE and CT states in Y6 (Y6Se) dimers.

\begin{table}[t]
\begin{ruledtabular}
\begin{tabular}{l c c c c c }
    Molecule & Singlet state & Triplet state & $L_{nm}$ & SOCME (cm-1)\\
    \colrule 
    Monomer & S1 & T1  & 0.9443 \\
    Monomer & S1 & T2 & 0.9106  \\
    \colrule
    D1 & CT(S) [S$_2$] & FE(T) [T$_1$] & 0.0055 & 0.19 \\
       & CT(S) [S$_2$] & FE(T) [T$_2$] & 0.0063 & -\\
       & CT(S) [S$_2$] & FE(T) [T$_3$] & 0.0087 & -\\
       & CT(S) [S$_2$] & FE(T) [T$_4$] & 0.0019 & -\\
       & CT(S) [S$_2$] & CT(T) [T$_5$] & 0.9685 & 0.01\\
       & CT(S) [S$_2$] & CT(T) [T$_6$] & 0.8100 & -\\
    \colrule
    D1 & FE(S) [S$_4$] & FE(T) [T$_1$] & 0.1199 & -\\
       & FE(S) [S$_4$] & FE(T) [T$_2$] & 0.8444 & 0.05\\
       & FE(S) [S$_4$] & FE(T) [T$_3$] & 0.0813 & -\\
       & FE(S) [S$_4$] & FE(T) [T$_4$] & 0.5599 & -\\
       & FE(S) [S$_4$] & CT(T) [T$_5$] & 0.0167 & 0.25\\
       & FE(S) [S$_4$] & CT(T) [T$_6$] & 0.0365 & - \\    
    \end{tabular}
    \end{ruledtabular}
    \caption{Spatial overlap metric $L_{nm}$ (Equation \ref{Lnmeq}) between different excited states for the Y6 monomer and D1 dimer. 
    The S$_2$ and S$_4$ states were chosen as by population analysis these have clear CT and FE character. 
    Transitions between distinct state characters (CT $\leftrightarrow$ FE) exhibit near-zero spatial overlap, and larger SOCMEs. 
    }

    \label{Lnmtab}
\end{table}

\subsection{HEOM dynamics}

\begin{figure*}
    \centering
    \includegraphics[width=1\linewidth]{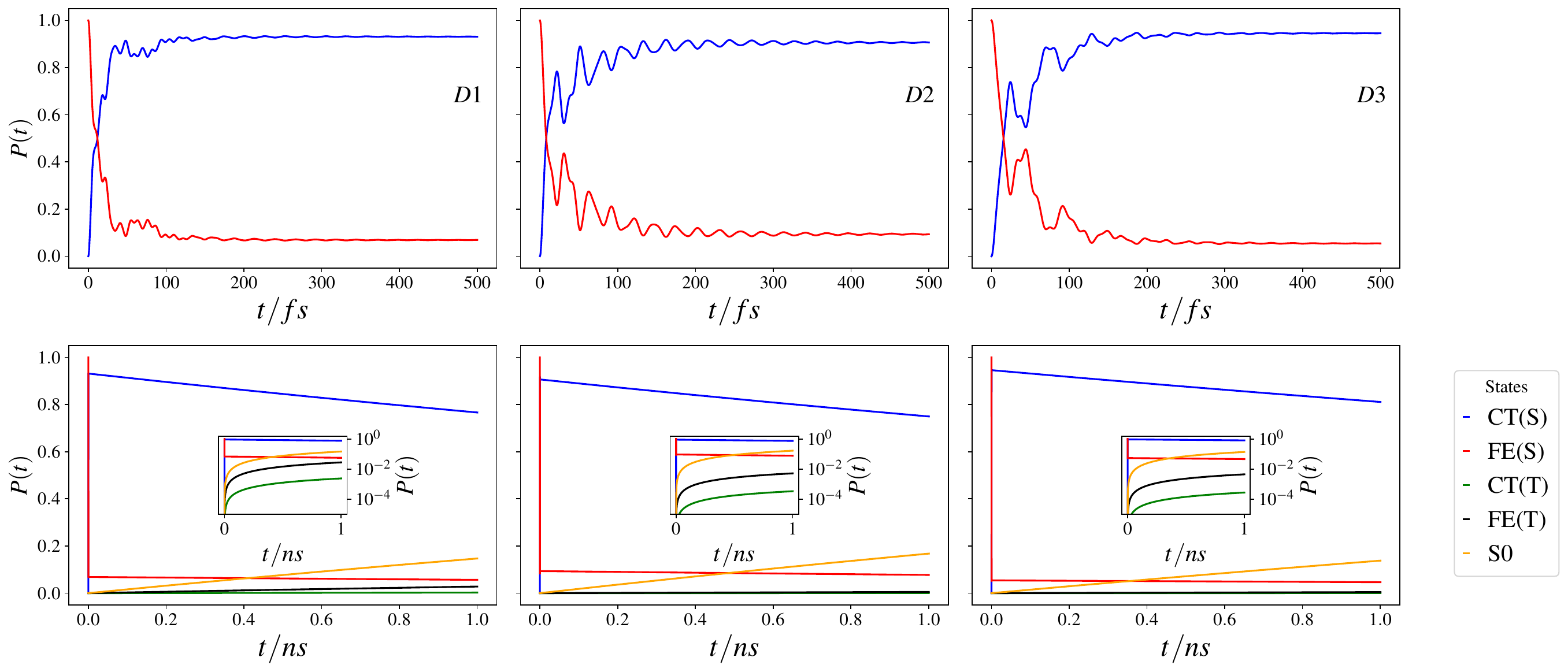}
    \caption{Population dynamics calculated using HEOM for the Y6 dimers (D1, D2, and D3) following occupation of the singlet FE state at $t=0$. 
    Top panels show the short-time non-adiabatic relaxation (${\cal{O}}(500\text{ fs})$); bottom panels show the long-time triplet formation (${\cal{O}}(1\text{ ns})$).
    State character are in the legend.}
    \label{HEOM_Y6}
\end{figure*}

\begin{figure*}
    \centering
    \includegraphics[width=1\linewidth]{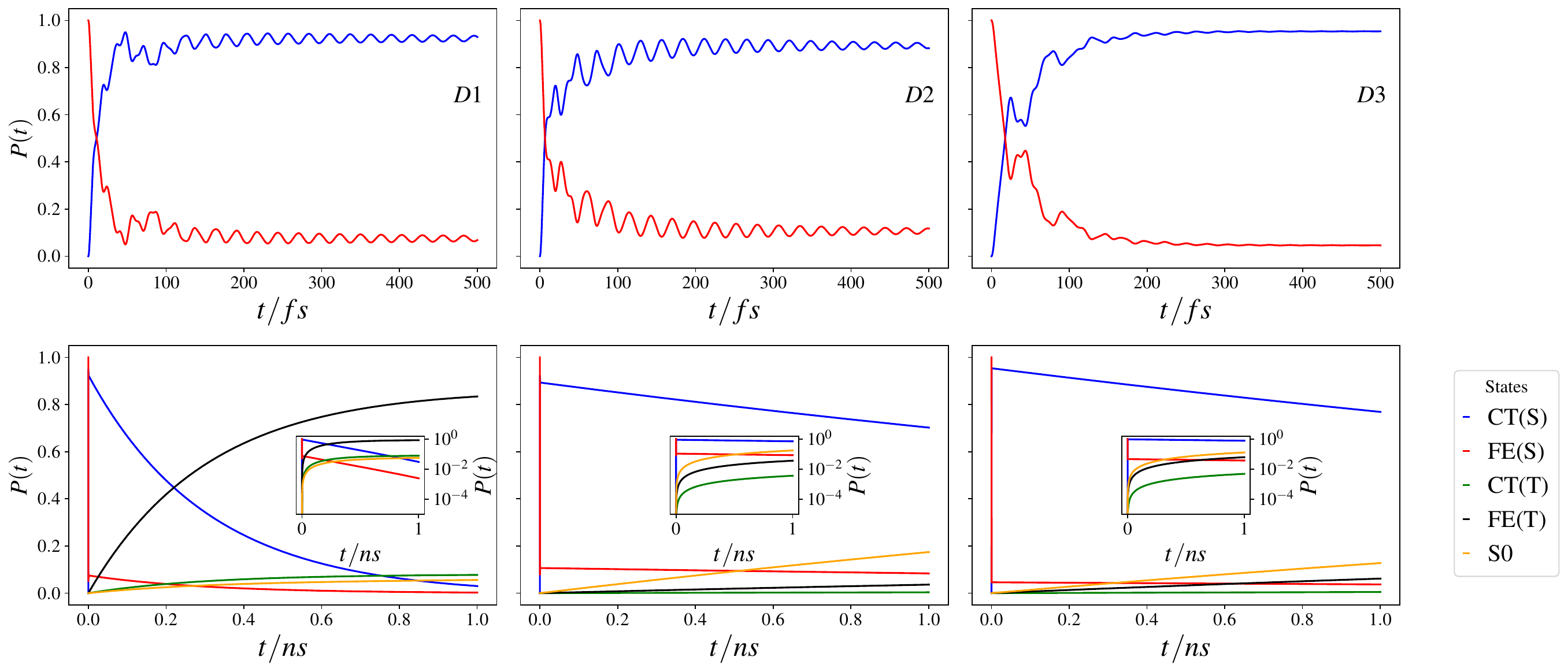}
    \caption{Population dynamics calculated using HEOM for selenised Y6Se dimers (D1, D2, and D3) following occupation of the singlet FE state at $t=0$. 
    Top panels show the short-time non-adiabatic relaxation (${\cal{O}}(500\text{ fs})$); bottom panels show the long-time triplet formation (${\cal{O}}(1\text{ ns})$).
    State character are in the legend.}
    \label{HEOM_Y6Se}
\end{figure*}

Following simulated photoexcitation (starting the system in a fully occupied singlet FE state at $t=0$), we propagate the exact non-Markovian HEOM dynamics for 1 ns.  
Figures \ref{HEOM_Y6} and \ref{HEOM_Y6Se} present the population evolution for representative aggregates (D1--D3; dynamics for D4-D6, which show similar behaviour, are provided in the Supplementary Information). 

In all configurations, population rapidly internally converts from the singlet FE to the singlet CT state within $\sim 100$ fs. 
On a nanosecond timescale, the triplet state is populated. 
This is in close agreement with estimates from transient absorption (TAS)\cite{Hart2023} and excited-state absorption (ESA)\cite{Mahadevan2024} measurements of solid-state films. 
Selenisation dramatically accelerates this process (Figure \ref{HEOM_Y6Se}), establishing significant triplet populations well within the 1 ns window, if there were competing decay processes we would therefore expect the Y6Se triplet yield to be significantly higher than Y6. 

Our key observation is that the triplet is generated via the singlet CT state, an intermolecular singlet state that can only exist in the solid state. 

\subsection{Effective rates and fluxes from HEOM}

\begin{figure*}
    \centering
    \includegraphics[width=1\linewidth]{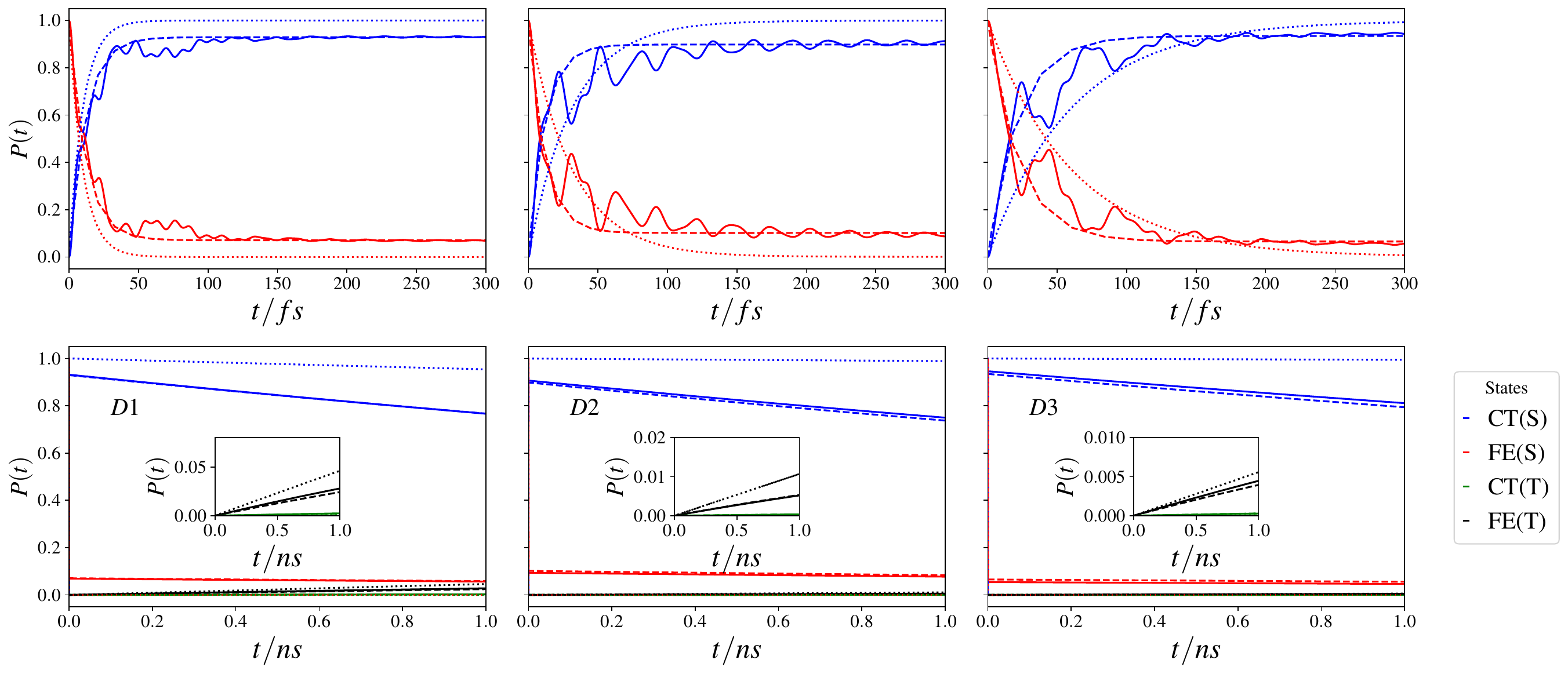}
    \caption{Figure to show the results for the dynamics obtained using HEOM (bold line), effective rates from HEOM (dashed line) and Marcus theory (dotted line)  for the Y6 dimers (shown left to right) D1, D2 and D3 following photoexcitation into the singlet FE state at $t=0$. The top panels show the short time dynamics (${\cal{O}}(300fs)$) and the bottom panels show the longer time dynamics (${\cal{O}}(1ns))$. The different types of state are as indicated in the legend.}
    \label{method_compare}
\end{figure*}

We extract time-independent effective rates from the HEOM memory kernel as described above, and plot the resulting Master equation with the actual HEOM data in Figure \ref{method_compare}. 
The effective rates show excellent agreement with the full simulation (dashed versus solid lines). 
This validates the approach, and justifies us using these rates to calculate the inter-state flux.

\begin{figure*}
    \centering
    \includegraphics[width=1\linewidth]{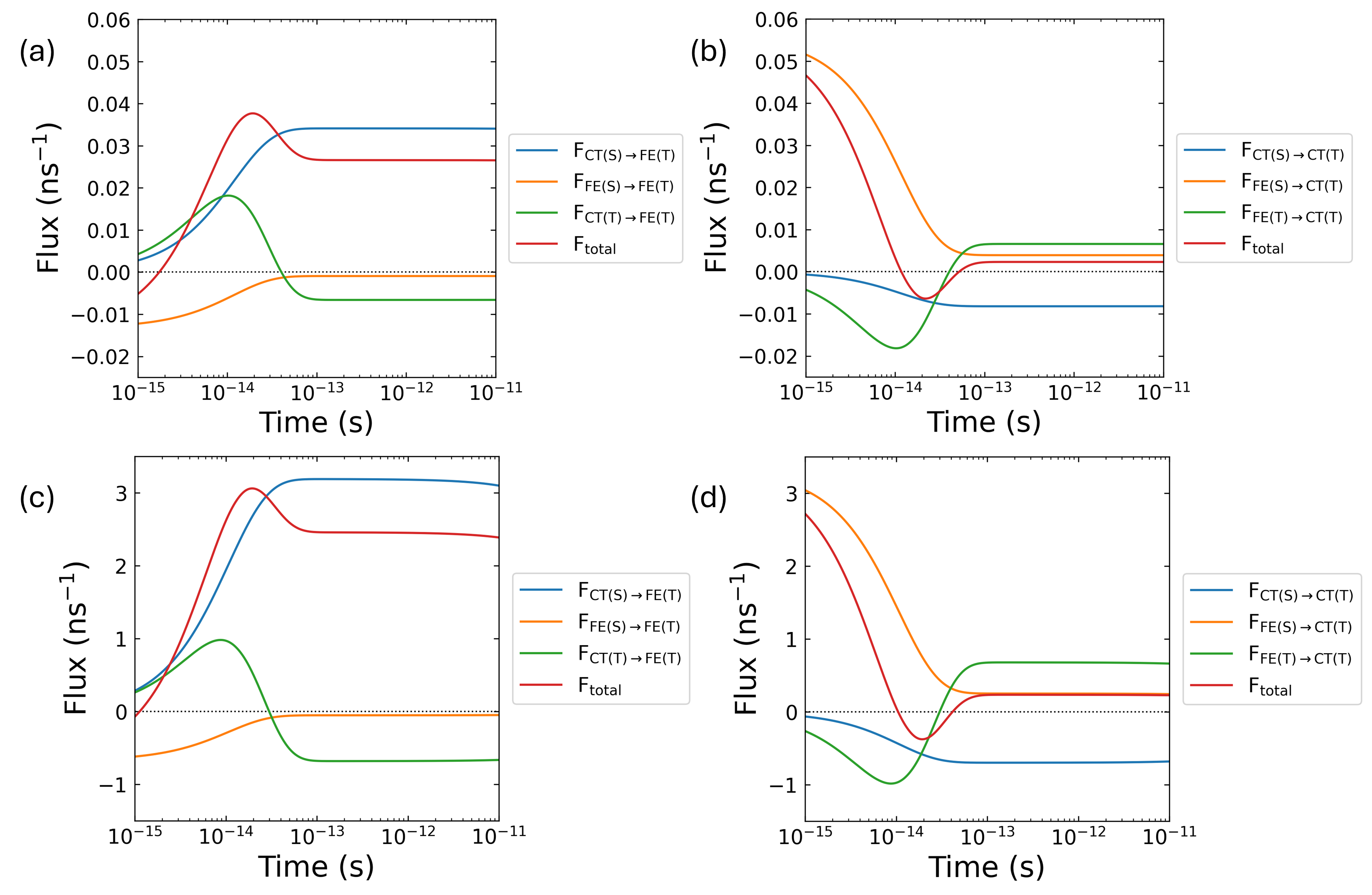}
    \caption{The net population fluxes into (a) the FE(T) state and (b) the CT(T) state for Y6 dimer D1. The bottom row shows the same figures, but calculated for Y6Se dimer D1. The different fluxes are as indicated in the legend, with $F_\mathrm{total}$ being the total net flux into the state.}
    \label{Fluxes}
\end{figure*}

Figure \ref{Fluxes} shows that the dominant triplet pathway is identical across both Y6 and Y6Se dimers: the transiently populated CT(S) state undergoes ISC into the FE(T) state (results for other Y6 and Y6Se dimers are qualitatively similar and shown in Figures \ref{Fluxes_Y6_D13}-\ref{Fluxes_YSe_D46} of the Supplementary Information). 
This is driven by the large CT(S) population generated from fast internal conversion from FE(S), and the presumably orbital symmetry allowed CT(S) $\leftrightarrow$ FE(T) spin-orbit coupling. 
A minor, direct, secondary pathway exists from $FE(S) \to CT(T)$. 
Selenisation preserves the same mechanism, but the overall flux is increased by almost an order of mangitude due to the heavy-atom amplified spin-orbit-coupling. 

Considering the CT(T) state, Figures \ref{Fluxes}b and \ref{Fluxes}d show that this populates mainly via the FE(S)-CT(T) pathway, again due to the large SOCME connecting these two states. 
For times $\gtrsim$ 100 fs, there is also a significant exchange of populations via the CT(S)-CT(T) and FE(T)-CT(T) pathways. 
The details of these processes differ between the dimers (see Figures \ref{Fluxes_Y6_D13}-\ref{Fluxes_YSe_D46}); perhaps these processes are not well described by our effective model, and we need more detail in our Hamiltonian. 

\begin{table}
\begin{ruledtabular}
\begin{tabular}{ c c c c c }
 Dimer & $k_\mathrm{CT(T), Y6}$ & $k_\mathrm{CT(T), Y6Se}$ & $k_\mathrm{FE(T),Y6}$ & $k_\mathrm{FE(T),Y6Se}$ \\
\colrule 
 D1 & $2.33\times10^6$ & $2.36\times10^8$ & $2.66\times10^7$ & $2.45\times 10^{9}$ \\ 
 D2 & $3.76\times10^5$ & $3.79 \times10^6$ & $5.87\times10^6$ & $3.68\times10^7$ \\ 
 D3 & $3.17\times10^5$ & $6.26\times10^6$ & $4.28\times10^6$ & $6.74\times10^7$ \\ 
 D4 & $1.09\times10^5$ & $4.23\times10^6$ & $6.41\times10^6$ & $1.76\times10^8$ \\ 
 D5 & $1.39\times10^5$ & $3.68\times10^7$ & $7.76\times10^6$ & $9.75 \times10^8$ \\ 
 D6 & $9.93\times10^4$ & $8.00 \times10^6$ & $4.51\times10^6$ & $2.65\times 10^8$ \\ 
\end{tabular}
\end{ruledtabular}
\caption{The initial rate of formation (in units of $\mathrm{s}^{-1}$) of the triplet FE and triplet CT states in the different Y6 and Y6Se dimers. For comparison the Y6 monomers rate of triplet formation without considering any vibrational effects is $9.2\times10^4~\mathrm{s}^{-1}$.}
\label{Initialrate_differenttypes}
\end{table}

By sampling these steady-state fluxes ($t>300$ fs), we extract a time-independent formation rate for the triplet species after initial internal conversion (Table \ref{Initialrate_differenttypes}). 
For all dimers, FE(T) formation outpaces CT(T) formation by an order of magnitude. 
These rates are a strong function of the dimer; the D1 H-aggregate, which exhibits the largest spatial overlap between units and the largest electron/hole transfer integrals, generates triplets an order of magnitude faster than the other packing motifs. 
Therefore we predict that the details of the micromorphology (crystal structure, degree of crystallanity) will have a very strong effect on the rate of formation of triplets, and therefore on upconversion device performance. 
Similarly, the propensity for Y6 to form triplets in organic photovoltaics (generally considered a loss mechanism) will vary considerably depending on the morphology and packing of the heterojunction. 

This extremely strong packing dependence is due to our central claim, that the main route for Y6 to form triplets is via an intermolecular singlet charge-transfer (CT(S)) state.

Finally, we assess the validity of Marcus theory. 
As shown in Figure \ref{method_compare} (dotted lines), Marcus theory successfully predicts the initial ($\sim 50$ fs) FE(S) $\to$ CT(S) transfer, and approximates the long-time detailed balance of FE(T). 
However, as a weak-coupling theory, it does not describe the early-time hybridisation of FE and CT manifolds, and therefore does not capture the hybridisation of the (CT and FE) singlet states at $\sim$ 100 fs. 
Consequently, it fails to predict the minor FE(S) $\to$ CT(T) pathway for generating triplets. 

Overall the HEOM-derived effective rates provide a quantitatively and qualitatively superior method for modelling long-time multi-state excited states in organic photovoltaic materials. 
Retaining Marcus theory rates in the workflow would enable validation of this approach in domains where a full HEOM simulation of dynamics was not computationally feasible.

\section{Conclusion and Future work}

We developed an \textit{effective} five-state model for Y6 and Y6Se dimers, as a surrogate for the solid state. 
The key observation from our Hamiltonian (Figure \ref{ham}) is that the spin-orbit coupling is relatively large for $S_{CT} \leftrightarrow T_{FE}$. 
By analysing the transition density matrix decomposed into fragments, made quantitative by a scalar similarity metric $L_{nm}$ (Table \ref{Lnmtab}), we see that this is associated with very different spatial locations of the FE versus CT excitations in Y6 (Y6Se) dimers. 
This macroscopic spatial symmetry breaking is, we assume, associated with a change in orbital symmetry, permitting spin-orbit-coupling (El-Sayed's rule). 
The spin-orbit coupling is thus an order of magnitude larger in the dimer than the monomer. 
The specific value is a strong function of the particular dimer, which we believe is due to this intramolecular charge-transfer state being very subtly dependent on the micromorphology\cite{Ward2026}. 
We propose this may have some relevance for the empirically observed sensitivity of Y6 to processing conditions. 

We then simulated the HEOM dynamics. 
We predict internal conversion from singlet FE to singlet CT on $\sim100$~fs for both Y6 and Y6Se. 
Triplets are then mainly produced via the CT(S)-FE(T) pathway, on a $\sim1$~ps timescale consistent with TAS\cite{Hart2023} and ESA\cite{Mahadevan2024} measurements. 
Souza et al.\cite{Souza2025} achieve a similar agreement based on small spin-orbit-couplings calculated in the monomer, but by relying on extreme intramolecular torsions (to \ang{90}) in the excited state to break the symmetry. 
We believe that in the solid-state such torsions would be sterically hindered. 

At longer times, on $\sim 1$~ns the model predicts that Y6 dimers generate triplets.  
These are a similar time scale to that believed to be observed in TAS spectroscopy for Y6. 

A Master equation rate model constructed with Marcus theory described most of the dynamics seen in HEOM, but missed the minor FE(S)-CT(T) route as it cannot describe the quantum recurrence and coherence, but was not therefore quantitative (i.e. correct yields) in the long time limit.

We applied the recent technique developed by Gestsson et al.\cite{Gestsson2025-ez} and used Nakajima-Zwanzig projection to calculate effective classical rates directly from the HEOM memory kernel. 
Though this integrated out the bath driven oscillations, it otherwise recovers the exact (non-perturbative, including quantum recurrences and coherences) dynamics, including quantitative yields, at a fraction of the computational cost of the full HEOM dynamics. 
Given the additional challenges we found converging the projection operator, on top of the validation required to converge the HEOM dynamics, we would suggest retaining Marcus modelling in most situations as a cross validation, particularly where the HEOM dynamics are too expensive to study the properties of interest directly. 

In immediate future work we intend to increase the size of our Hamiltonian---we believe that our inability to see a clear trend between dimers in the CT(S)-CT(T) and FE(T)-CT(T) was due to us having combined multiple distinct processes in our effective states. 
These larger Hamiltonians are likely to require GPU acceleration in the current code\cite{Huang2023}, or moving to a tensor network HEOM code\cite{Ke2023}.

Our proposal of fast intersystem crossing being driven by an an implicitly \textit{intramolecular} charge-transfer state stimulates a number of follow up investigations. 
As the transiently occupied charge transfer state is intermolecular, it is only present in dimers and larger solid-state aggregates. 
Therefore our mechanism would predict considerably faster triplet formation in the solid state than solution, whereas Souza et al.'s\cite{Souza2025} vibrational mechanism should be faster in the sterically unhindered liquid. 
First, this should be experimentally proved or disproved, perhaps by a repeat of the previous TAS\cite{Hart2023} or ESA\cite{Mahadevan2024} methods, but now looking at solution (or frozen solution) versus the solid state (including different deposition methods, post-deposition annealing etc.).

On the computational side we are stimulated to look at other classes of fused-ring electron-acceptors, such as the ITIC series, and see whether the same highly morphology dependent intersystem crossing route dominates. 
From a molecular design point of view, this realisation behoves us to model other heteroatom modifications to Y6, to see whether intersystem crossing can be further enhanced, or can be engineered with more stable and less challenging chemistry than selenium. 
More generally we will expand our model to look at dynamics in larger aggregates (such as calculating exciton diffusion rates), and modelling the full Y6--rubrene bilayer upconverting interface.

\section{Author contributions}

Contributor Role Taxonomy (CRediT). 
I.C.: 
Formal Analysis (lead); 
Investigation (lead); 
Methodology (equal);
Writing – original draft (equal); 
Writing – review and editing (equal).
L.H.: 
Investigating (supporting);
Methodology (equal);
Writing - original draft (supporting); 
P.V.:
Investigation (supporting);
Methodology (supporting);
T.W.: 
Investigating (supporting);
J.M.F.: 
Conceptualization (lead); 
Methodology (equal);
Supervision (lead);
Writing – original draft (equal); 
Writing – review and editing (lead).

Quantum Chemistry calculations, and nonadiabatic dynamic input files and intermediate files are available online\cite{Figshare}. 

\section{Acknowledgement}
We are grateful for fruitful discussion with Jenny Nelson, Hanbo Yang, Tim Rein and Emily Wentworth. 
The key approach of extracting effective rates from HEOM was suggested by Hallman Gestsson and Alexandra Olaya-Castro\cite{Gestsson2025-ez}. 

J.M.F. 
is supported by a Royal Society University Research Fellowship
(URF-R1-191292). 
T.W. is a Royal Society funded PhD student on the above grant (URF-R1-191292). 
I.C. is supported by EPSRC (EP/Y020790/1).
L.J.F.H. is supported by an EPSRC Doctoral Prize Fellowship and the UKRI via the ERC underwrite scheme (EP/Z533361/1). 

We gratefully acknowledge the use of the Imperial College Research Computing Service\cite{HPC}.
Additionally, via our membership of the UK's HEC Materials Chemistry Consortium (MCC), funded by EPSRC (EP/R029431 and EP/X035859), this work used the \textsc{ARCHER2} UK National Supercomputing Service (http://www.archer2.ac.uk).

\section{References}
\bibliography{NonadiabaticDynamicsUpconversion}

\clearpage 
\onecolumngrid

\setcounter{equation}{0}
\setcounter{figure}{0}
 \setcounter{section}{0}
\setcounter{table}{0}
\setcounter{page}{1}
\makeatletter
\renewcommand{\theequation}{S\arabic{equation}}
\renewcommand{\thefigure}{S\arabic{figure}}
\renewcommand{\thetable}{S\arabic{table}}
\renewcommand{\thesection}{S\arabic{section}}
\renewcommand{\thepage}{S\arabic{page}}
\renewcommand{\bibnumfmt}[1]{[S#1]} 
\renewcommand{\citenumfont}[1]{S#1} 

\section*{Supporting Information}

\section{Paramaterising the effective Hamiltonian}

For Y6 monomers, the three lowest energy states are the singlet S1 state followed by two lower in energy triplet states T1 and T2.  We therefore anticipate that to initially understand the dynamics of the Y6 dimers after excitation, there are six relevant triplet states (4 FE states and 2 CT states) and 4 singlet states (2 FE states and 2 CT states).  We initially calculate the 10 lowest energy triplet and singlet states, and find that the relevant states are the six lowest energy triplet and four lowest energy singlet states which make up the visible spectrum energy window in the dimer. 

From a transition density matrix population analysis in TheoDORE\cite{plasser2020theodore} with the molecules as the two sites, we ascribe the charge-transfer nature of the states. 
Following Zhang et al.\cite{zheng2017effect}, we classify `CT' states as those states with $\omega_{CT} \gtrsim 0.75$ and FE states as those states with $\omega_{CT} \lesssim 0.25$, and declaring intermediate states as `mixed'. 

For the Y6 and Y6Se dimers considered, we find that the two lowest in energy singlet states are CT states followed by two higher in energy FE states. By contrast, for the triplet states, we find that the lowest four in energy excited states are FE states followed by two higher in energy CT states.  In section \ref{theodore} (Figures. \ref{CT_states} and \ref{E_states}) we show this analysis of the charge transfer character and the energies of the different excited states for all 6  dimers of Y6 and Y6Se considered in this work.  
As the motivation of selenising Y6 is that Se (Z=34) has larger relativistic spin-orbit-coupling than S (Z=16), one should check whether the underlying electronic structure (rather than just the spin-orbit-coupling) is affected. 
In section \ref{rel_Se} (Figures. \ref{Y6Serel1}, \ref{Y6Serel2}, \ref{Y6Serel3} and \ref{Y6Serel4})  we show that relativistic corrections to the site energies \cite{vanLenthe1996} and state properties are  negligible for Y6Se, and are not considered in the construction of the matrix elements in this work.  

Each of the dimers considered has two singlet CT states, two singlet FE states; and four triplet FE states and two CT triplet states. 
From these states, we can define one effective excited state for each different type of state, i.e, an effective CT singlet state, an effective FE singlet state, an effective CT triplet state and an effective FE triplet state.

The effective site energy are the arithmetic mean of the TDDFT energies, i.e. 
\begin{equation}
    E_{S_{FE}}=\frac{1}{2}\sum_{n=1}^2 E_{FE_n, S} \leftrightarrow  \quad    E_{T_{FE}}=\frac{1}{4}\sum_{n=1}^4 E_{FE_n, T}, 
    \label{effective_e1}
\end{equation}
\begin{equation}
    E_{S_{CT}}=\frac{1}{2}\sum_{n=1}^2 E_{CT_n, S} \quad     E_{T_{CT}}=\frac{1}{2}\sum_{n=1}^2 E_{CT_n, T} .
    \label{effective_e2}
\end{equation}

We can then obtain the coupling matrix elements by taking the quadrature sum over the coupling between the individual states, with 
 \begin{equation}
        V_{S_{FE}, T_{FE}} = \sqrt{\sum_{n=1}^2 \sum_{m=1}^4 SOCME_{FE_n(S), FE_m(T)}^2}, 
        \label{coupling1}
        \end{equation}
       \begin{equation} 
        V_{S_{CT}, T_{CT}} = \sqrt{\sum_{n=1}^2 \sum_{m=1}^2 SOCME_{CT_n(S), CT_m(T)}^2}, 
    \end{equation}
    \begin{equation}
        V_{S_{FE}, T_{CT}} =\sqrt{ \sum_{n=1}^2 \sum_{m=1}^2 SOCME_{FE_n(S), CT_m(T)}^2}, 
        \end{equation}
\begin{equation}
        V_{S_{CT}, T_{FE}} = \sqrt{\sum_{n=1}^2 \sum_{m=1}^4 SOCME_{CT_n(S), FE_m(T)}^2 } , 
    \end{equation}
    \begin{equation}
  V_{S_{FE}, S_{CT}} = V_{T_{FE}, T_{CT}}= \sqrt{2( V_h^2+V_e^2)}, 
    \end{equation}
    \begin{equation}
          V_{S_0 ; T_{FE}}=\sqrt{\sum_{n=1}^4 SOCME_{S_0, FE_n(T)}^2},
          \end{equation}
and 
          \begin{equation}
               V_{S_0 ; T_{CT}}=\sqrt{\sum_{n=1}^2 SOCME_{S_0, CT_n(T)}^2}. 
               \label{coupling2}
          \end{equation}

Here, $V_h$ and $V_e$ are the hole and electron transfer integrals calculated using the group's counterpoise method \cite{Kirkpatrick2007, Baumeier2010} and $SOCME$ are the SOCME between various states in the dimer using Spin–Orbit Mean Field theory \cite{He1996} method in ORCA 6 using TDDFT B3LYP and 6-31G(d, p) basis.  We choose to set both $V_{S0; FE_{eff},S}$ and $ V_{S0; CT_{eff},S}$  to zero due to large energy gap between the ground state and the singlet excited states \cite{Englman1970}.  For simplicity, in this work,  we will drop the effective label when discussing the different states.

\section{Y6 and Y6Se dimer structures}

\begin{figure*}[htbp]
    \centering
    
    \begin{overpic}[width=0.48\textwidth]{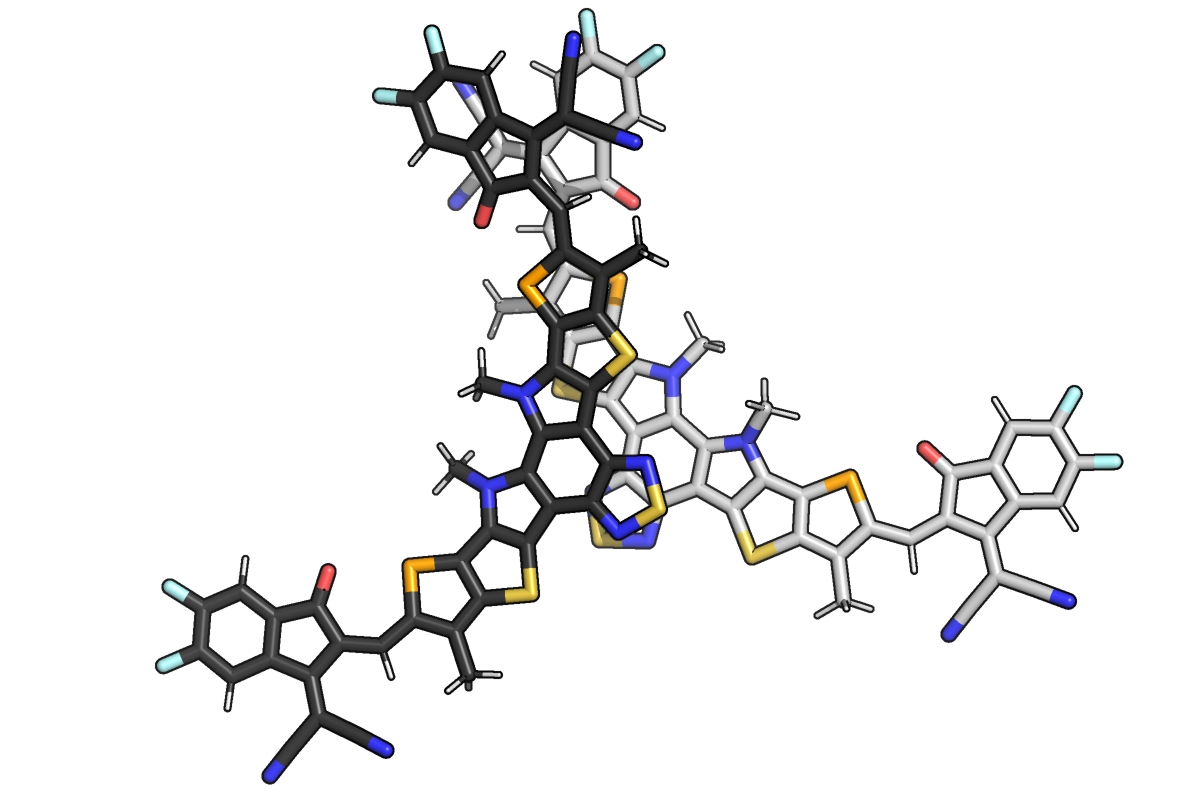}
        \put(5,5){\Large \textbf{D1}}
    \end{overpic}
    \hfill
    \begin{overpic}[width=0.48\textwidth]{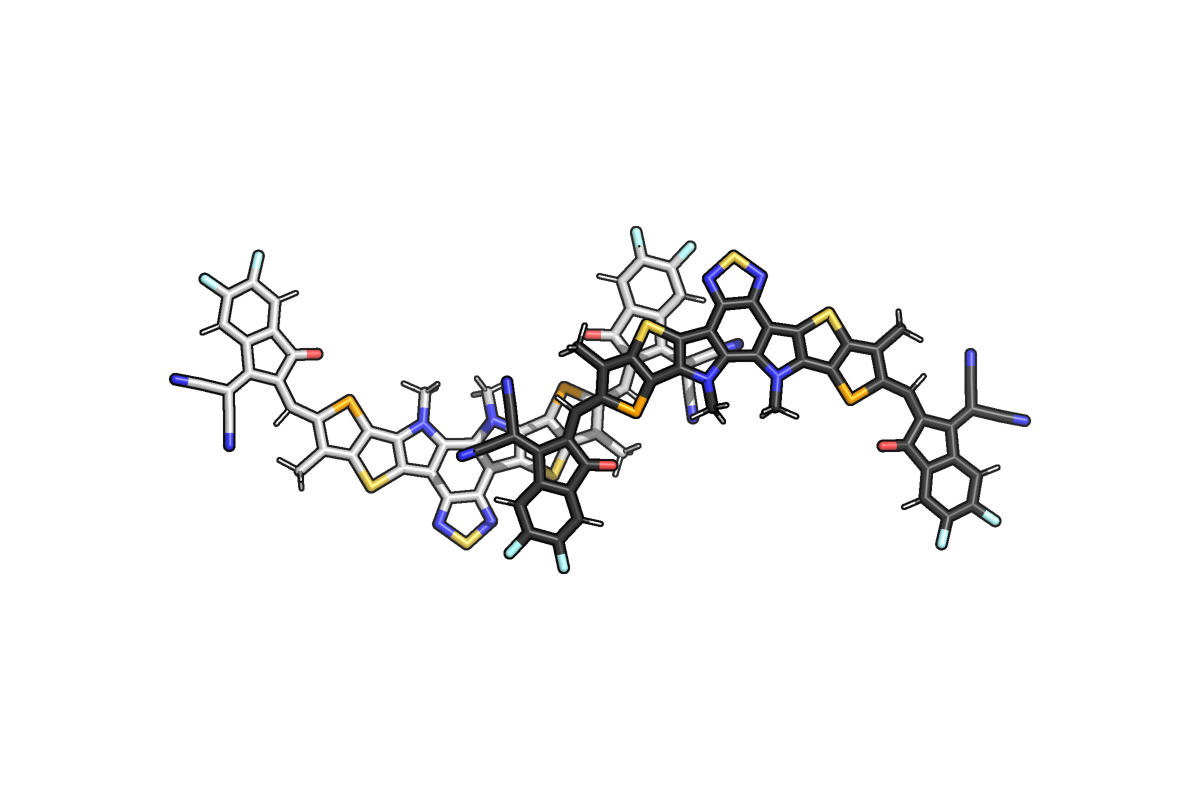}
        \put(5,5){\Large \textbf{D2}}
    \end{overpic}
    
    
    \begin{overpic}[width=0.48\textwidth]{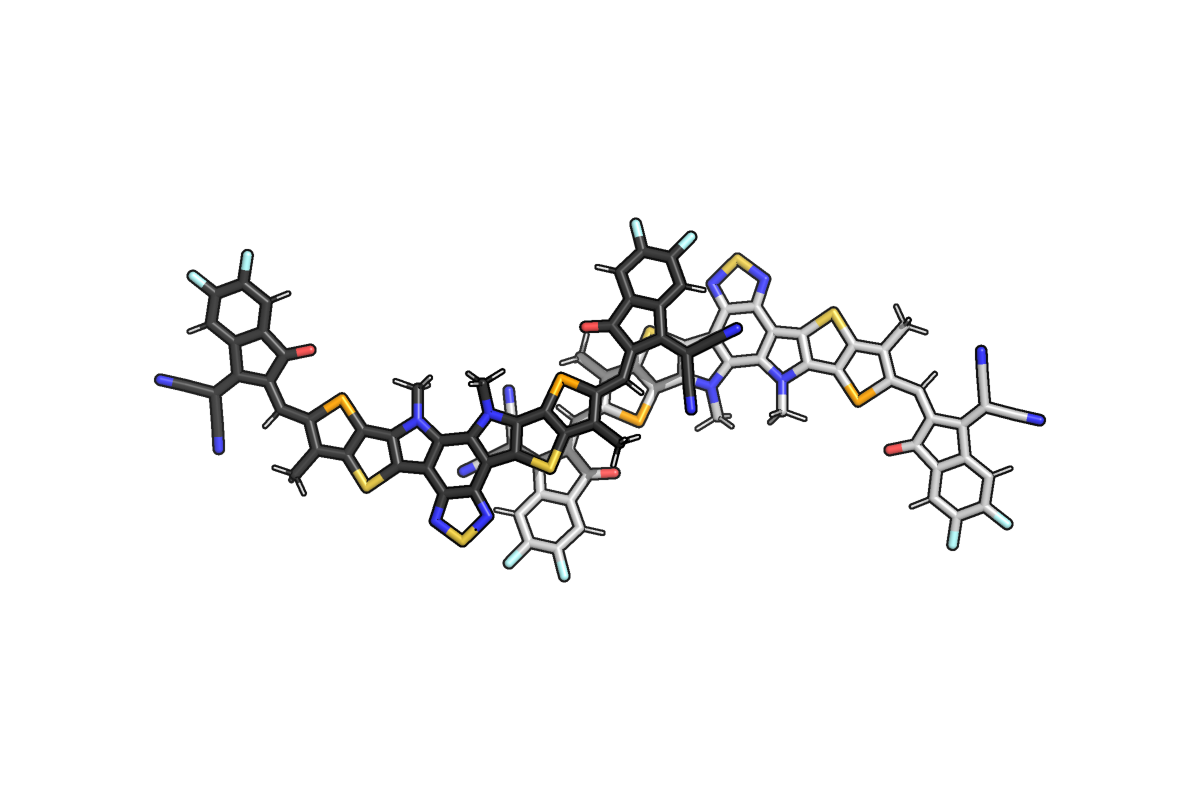}
        \put(5,5){\Large \textbf{D3}}
    \end{overpic}
    \hfill
    \begin{overpic}[width=0.48\textwidth]{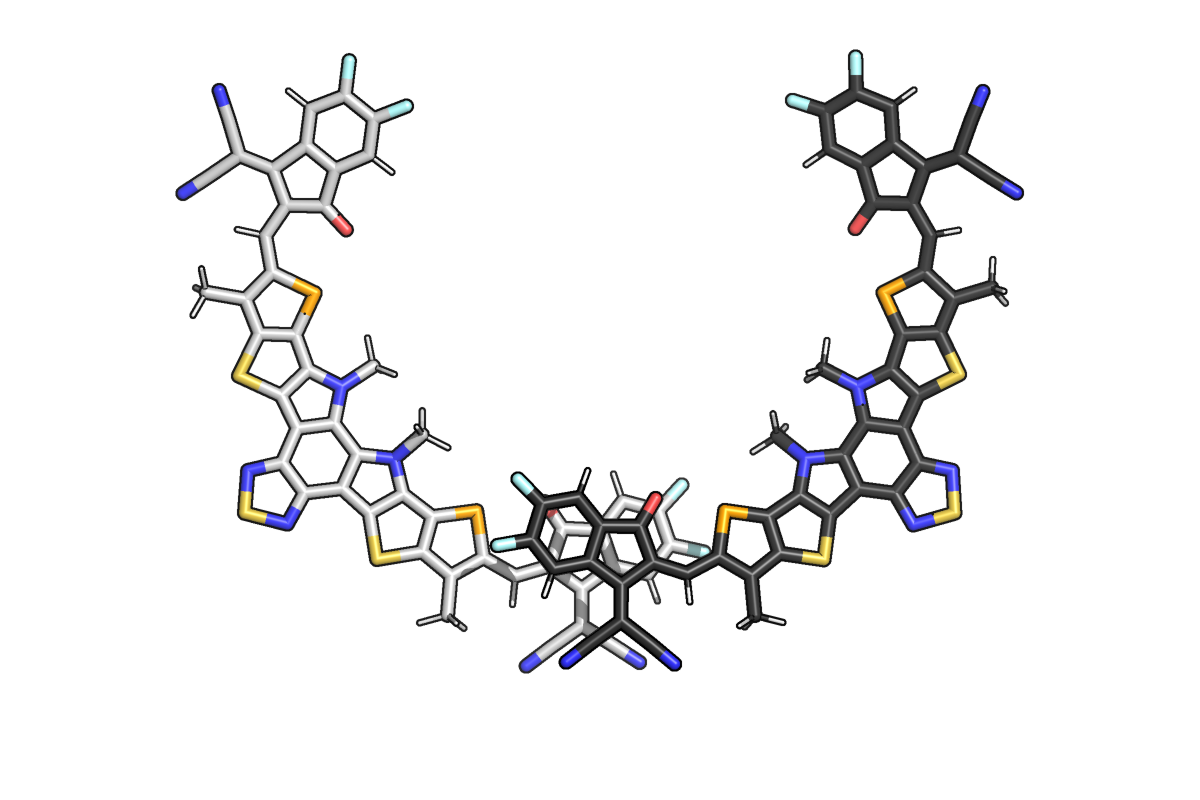}
        \put(5,5){\Large \textbf{D4}}
    \end{overpic}
    
    
    \begin{overpic}[width=0.48\textwidth]{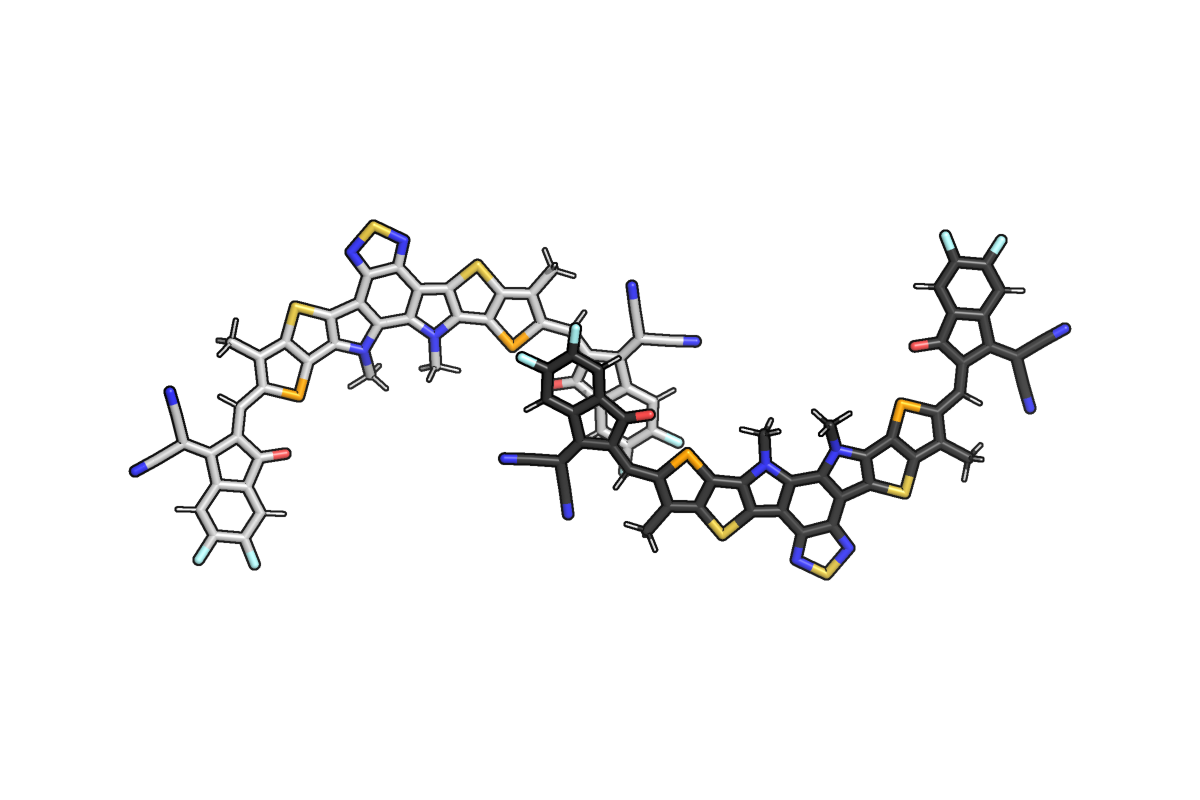}
        \put(5,5){\Large \textbf{D5}}
    \end{overpic}
    \hfill
    \begin{overpic}[width=0.48\textwidth]{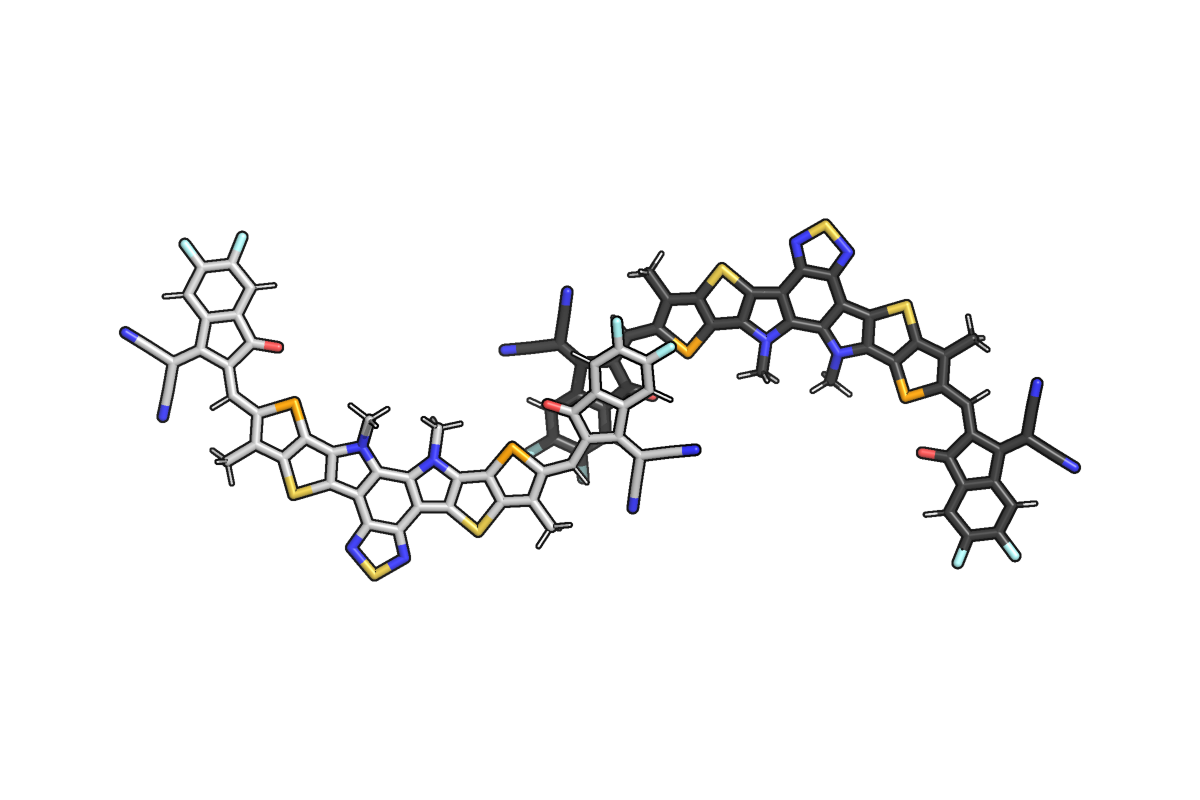}
        \put(5,5){\Large \textbf{D6}}
    \end{overpic}
    
    \caption{The six distinct dimer configurations. The structures were originally from a recent high-quality Y6 crystal structure solution\cite{Xiao2020}, then post-processed and relaxed by hybrid DFT by Giannini et~al.\cite{giannini2024role}. 
    The Y6Se structures are generated directly from the Y6 structures (with no additional relaxation), with the Se substitution for the two external sulphurs, as shown above in dark-orange.}
    \label{fig:dimer_structures}
\end{figure*}

Figures \ref{fig:dimer_structures} show the structures of the D1-D6 dimers of Y6/YSe. 

\section{Properties of Singlet and Triplet states}\label{theodore}

\begin{figure}
    \centering
    \includegraphics[width=1\linewidth]{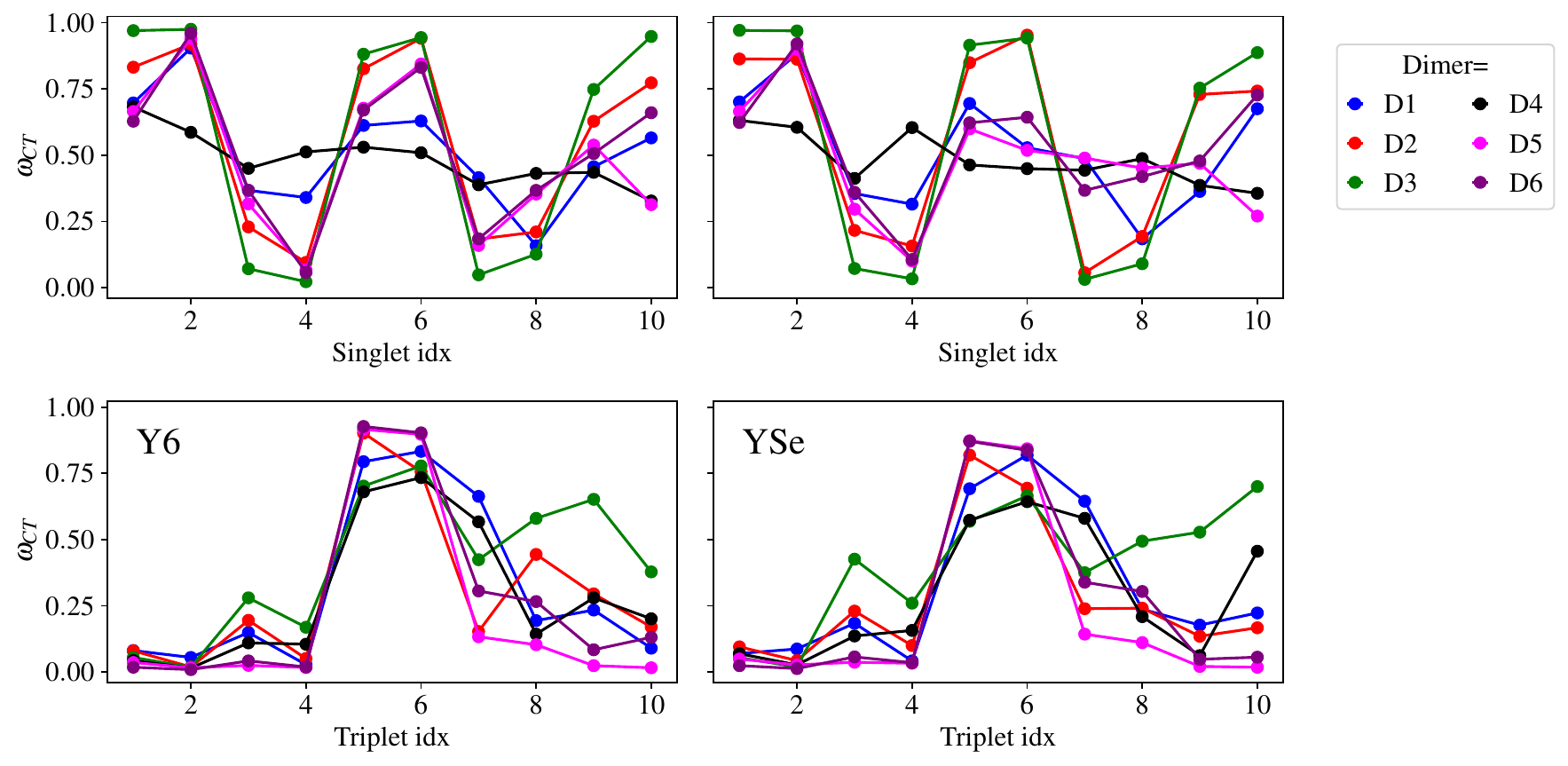}
    \caption{Figure to show the extent of charge transfer (probed by $\omega_{CT}$) of the singlet and triplet states of the different Y6 and Y6Se dimers.}
    \label{CT_states}
\end{figure}

\begin{figure}
    \centering
    \includegraphics[width=1\linewidth]{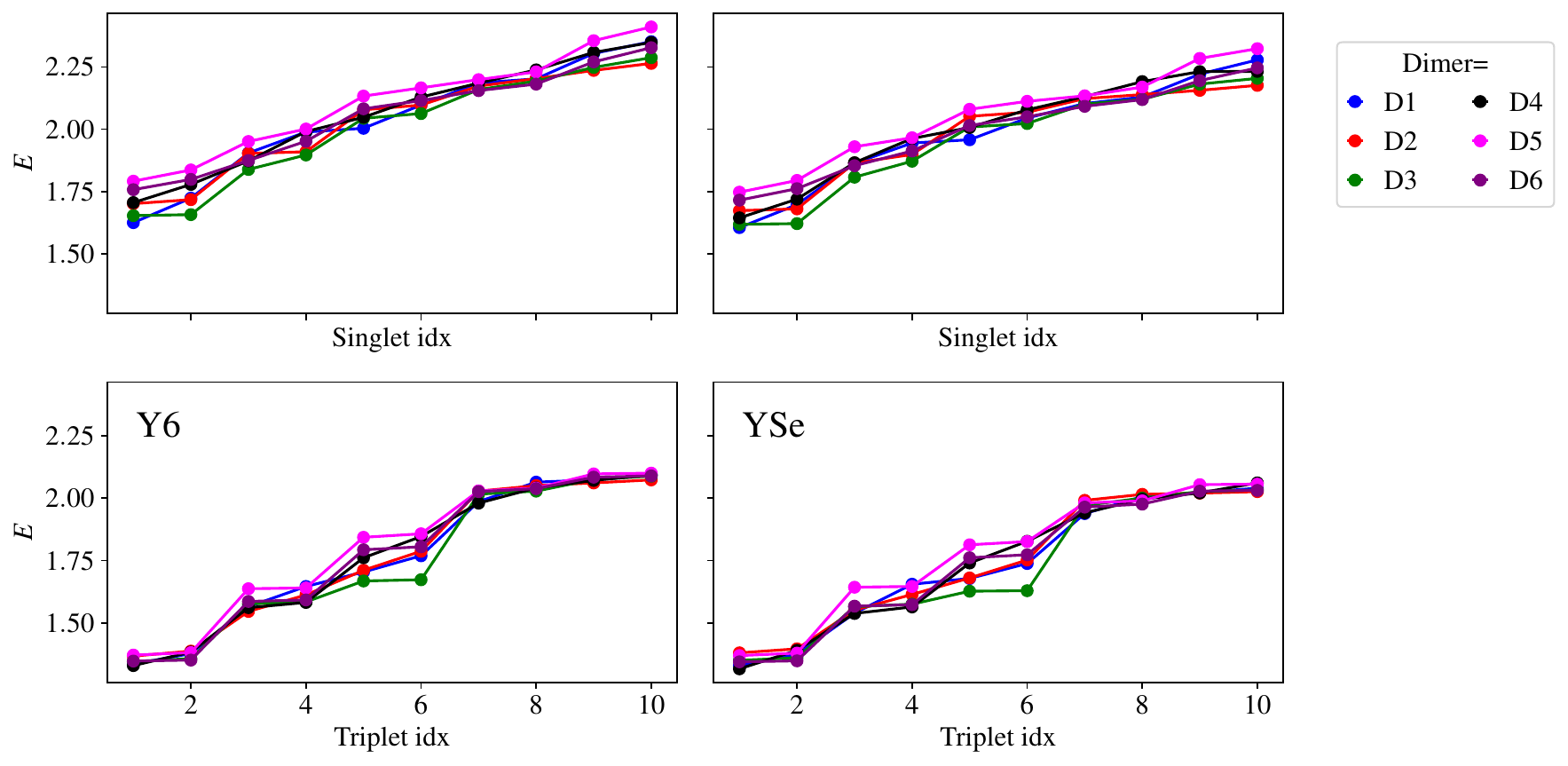}
    \caption{Figure to show the energy ($E$ in eV) of the singlet and triplet states of the different Y6 and Y6Se dimers.}
    \label{E_states}
\end{figure}

Figures \ref{CT_states} and \ref{E_states} show the analysis of the charge transfer character ($\omega_{CT}$) and energy ($E$) of the different Y6 and Y6Se dimers respectively. 

\section{Relativistic electronic structure for Y6Se} \label{rel_Se}

In order to validate the electronic structure method used for Y6Se, since Se is slightly further down the periodic table we considered the effects of relativity on the the singlet and triplet states of the different dimers of Y6Se.

\begin{figure}
    \centering
    \includegraphics[width=1\linewidth]{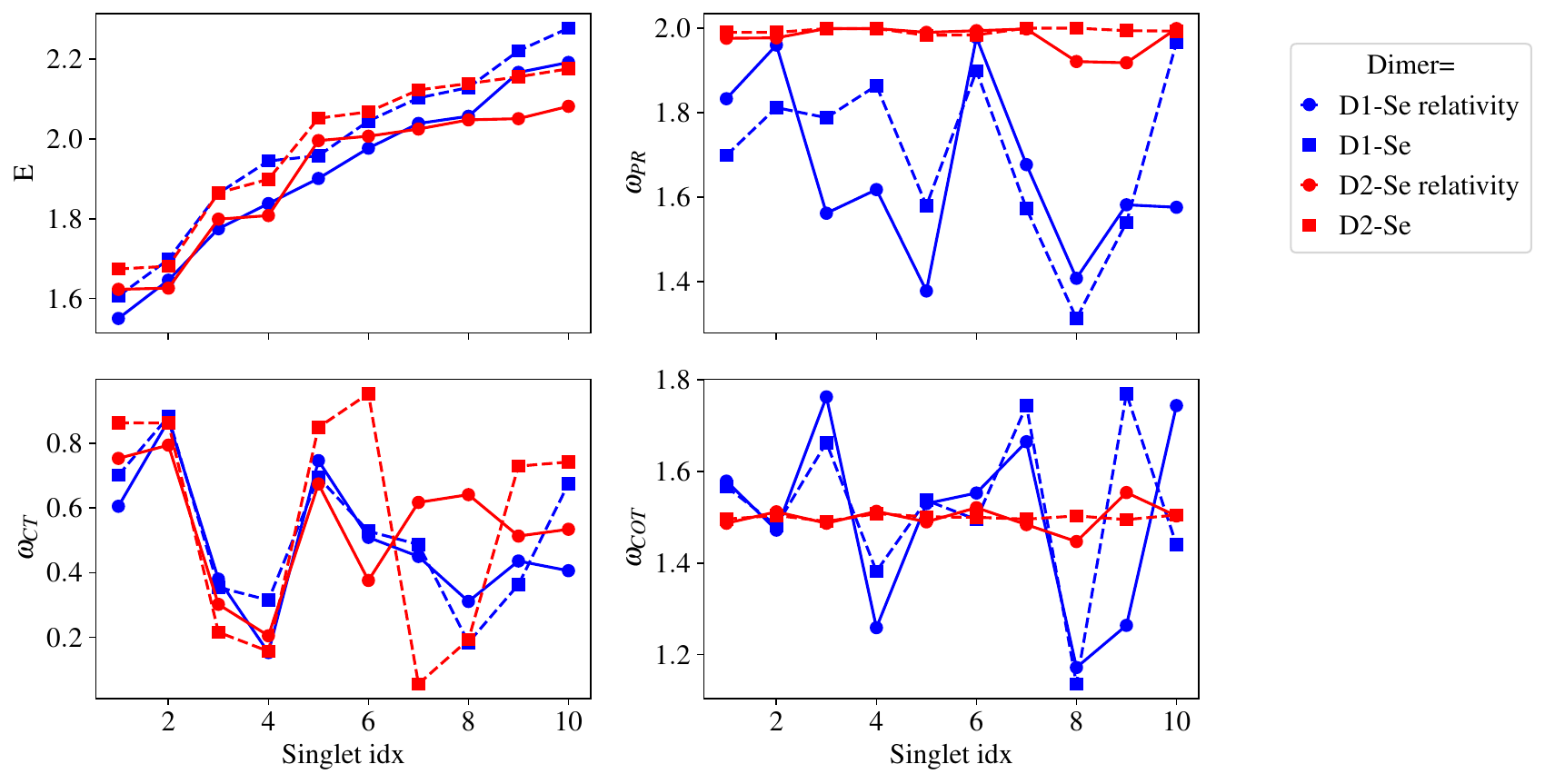}
    \caption{How the energy ($E$) and the extent of charge transfer ($\omega_{CT}$) of the Y6Se D1 and D2 dimers singlet states  calculated with relativistic corrections compare to those calculated without relativistic corrections.We here consider the D1 and D2 dimers, similar results are seen for other dimers.  }
    \label{Y6Serel1}
\end{figure}

\begin{figure}
    \centering
    \includegraphics[width=1\linewidth]{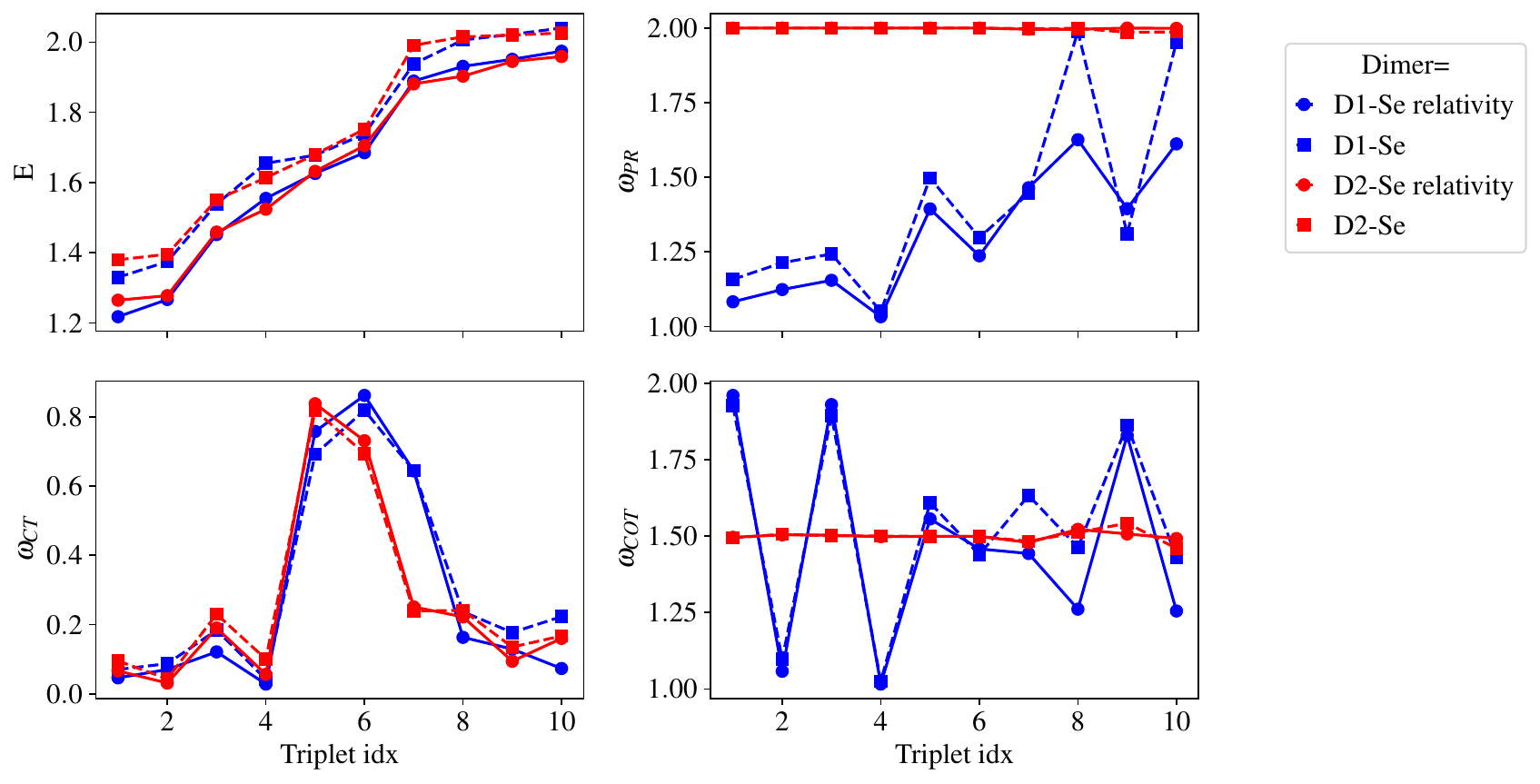}
    \caption{How the energy ($E$) and the extent of charge transfer ($\omega_{CT}$) of the Y6Se D1 and D2 dimers triplet states  calculated with relativistic corrections compare to those calculated without relativistic corrections. We here consider the D1 and D2 dimers, similar results are seen for other dimers.}
    \label{Y6Serel2}
\end{figure}

In figures \ref{Y6Serel1} and \ref{Y6Serel2} we show that the state properties calculated using relativistic corrections (ZORA calculations \cite{Neese2011} in Orca) are very similar to those calculated without relativistic corrections, but that there are modest shifts in the site energies from our TDDFT B3LYP calculations to the ZORA calculations.

 In order to understand the slight shift in site energies when relativity was considered we then took a deeper dive into the calculations. In order to run the relativistic calculations in ORCA the basis set has to be changed and the TDA approximation - known to get accurate singlet and triplet states \cite{Woon2022} - must be turned off.

\begin{figure}
    \centering
    \includegraphics[width=1\linewidth]{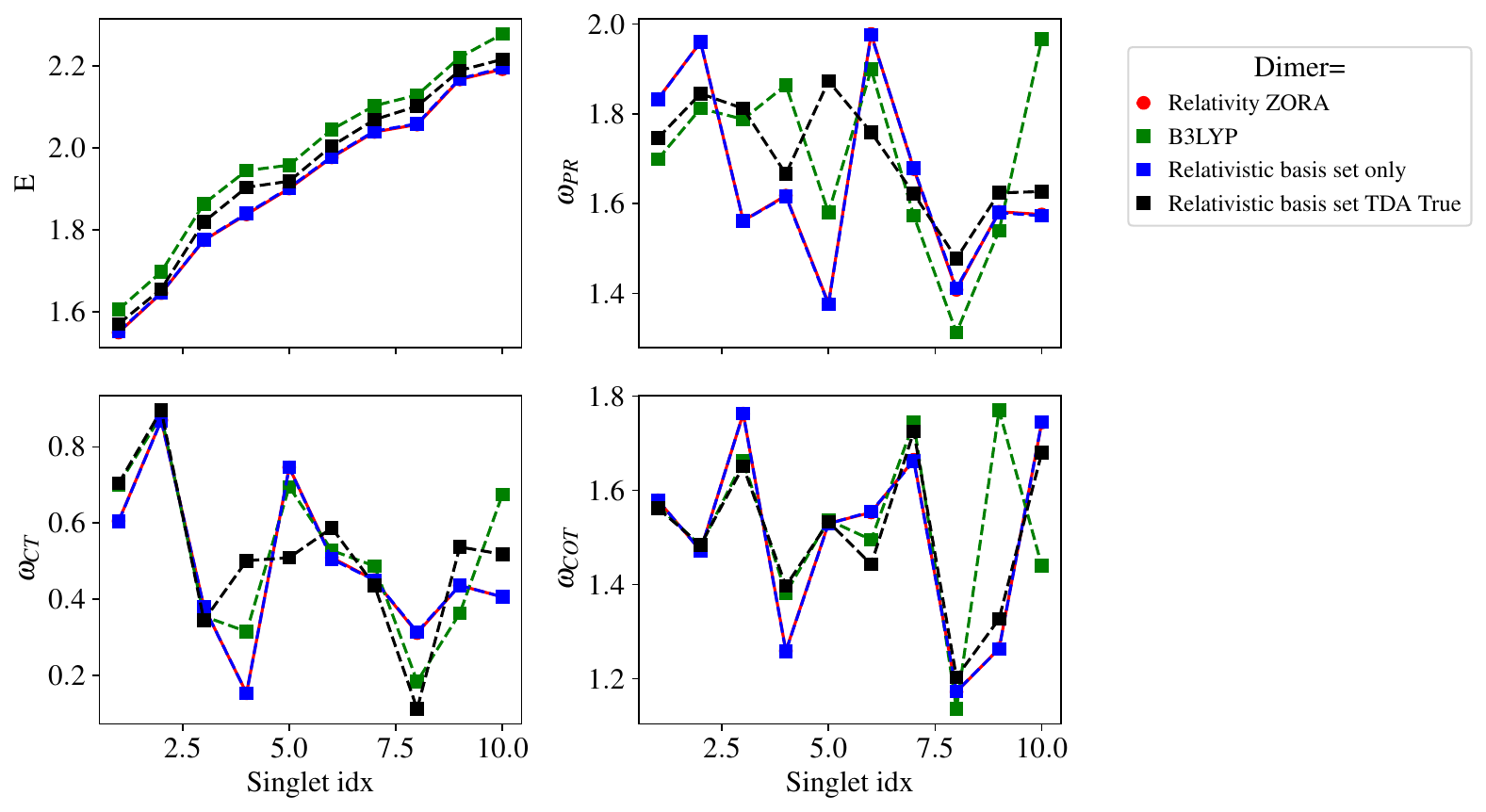}
    \caption{How the change of basis and turning off the TDA approximation affects the energy ($E$) and extent of charge transfer ($\omega_{CT}$) of the Y6Se D1 dimer singlet excited states compared to relativistic ZORA calculation and `standard' TDDFT calculations. We here examine the D1 dimer with essentially identical  results being seen for other dimers.}
    \label{Y6Serel3}
\end{figure}

\begin{figure}
    \centering
    \includegraphics[width=1\linewidth]{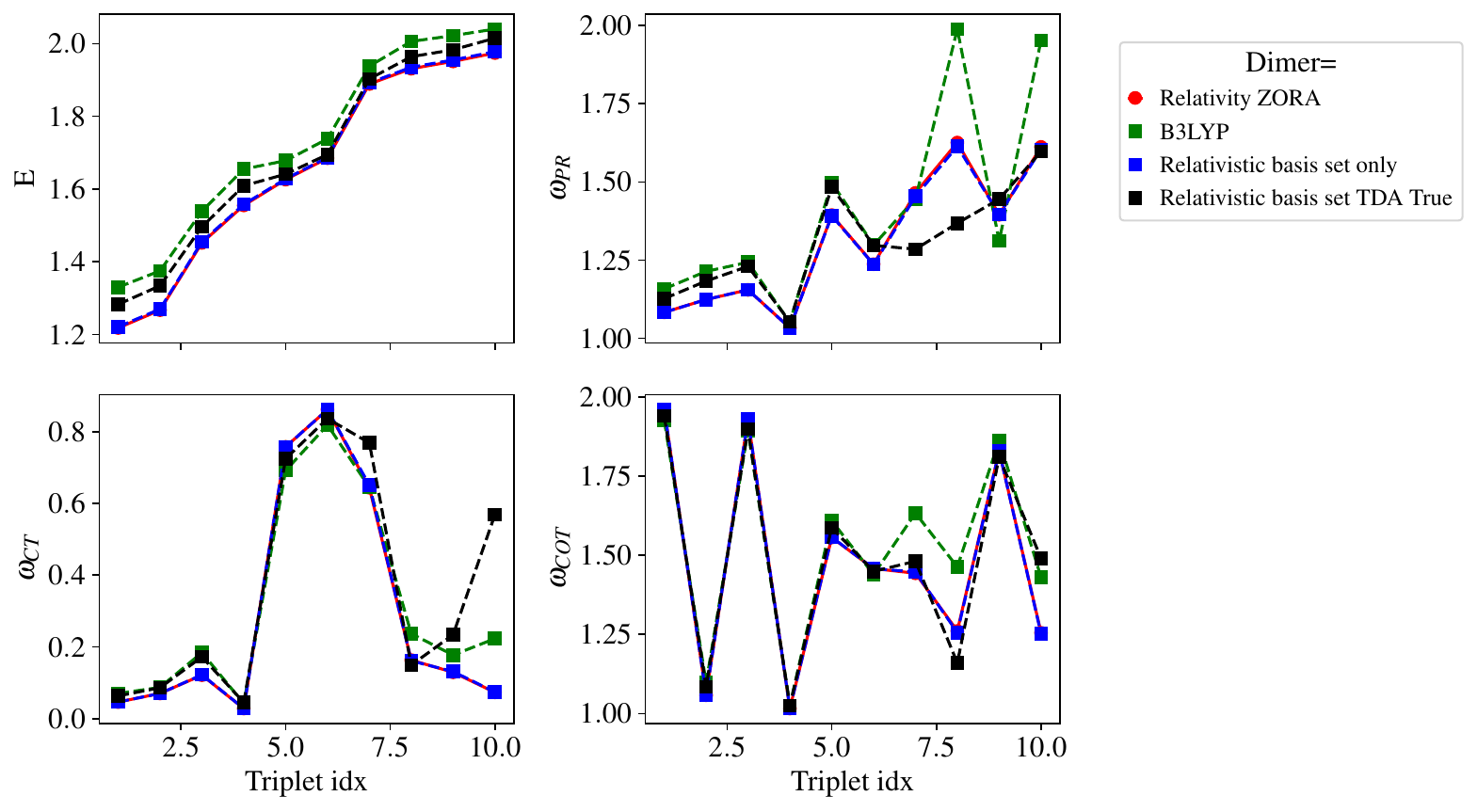}
    \caption{How the change of basis and turning off the TDA approximation affects the energy ($E$) and extent of charge transfer ($\omega_{CT}$) of the Y6Se D1 dimer triplet excited states compared to relativistic ZORA calculation and `standard' TDDFT calculations. We here examine the D1 dimer with essentially identical  results being seen for other dimers.}
    \label{Y6Serel4}
\end{figure}

Figures \ref{Y6Serel3} and \ref{Y6Serel4} show that the change in energies of the excited states from the relativistic calculations can mainly be attributed to turning off the TDA approximation rather than any relativistic effects. In the main text, we therefore run Y6Se matrix elements using the standard B3LYP TDDFT approach.

\section{HEOM details}

HEOM `exactly' solves the dynamics of a quantum system in contact with a Bosonic environment (the Bath) given by: 
\begin{equation}
\hat H = \hat H_S + \hat H_B + \hat H_{SB} . 
\label{Hamil_eq}
\end{equation}

Here the System $\hat H_S$ is our five-state $\hat H_{\text{eff}}$ as shown in Figure \ref{ham} in the main body of the paper. 
To model the Bath, we assume that each state couples to an independent set of harmonic oscillators which represent molecular vibrations. 
We assume that each System state has the same coupling strength to the Bath modes. 
This means that the spectral density of the system takes the form 
\begin{equation}
J(\omega)=\sum_k g_k^2 \;  \delta(\omega-\omega_k), 
\end{equation} 
where $g_k$ and $\omega _k$ represent the strength of the electron-phonon coupling and the frequency of the $k^{th}$ Bath mode, respectively. 
Within HEOM, this discrete spectral density is replaced by continuous function to reflect the broadening of the phonon modes which occurs at finite temperature. 
From $J(\omega)$, one can calculate the bath correlation function, $C(t)$ 
\begin{equation}
C_k(t)=\frac{1}{\pi}\int_{-\infty}^\infty d\omega J_k(\omega)(1+n(\omega))e^{-i\omega t}
\end{equation}
where $n(\omega)$ is the Bose-Einstein distribution function. To approximate $C(t)$ when using HEOM, it is usual to do perform a Matsubara Expansion such that
\begin{equation}
    C_k(t) \simeq \sum_{k=1}^{K} c_{k} \exp[-\gamma_{m, k} t], 
\end{equation}
where  the infinite series is truncated after $K$ terms, our first convergence parameter and $\gamma_{m, k} \propto k_B T$. 

Within HEOM, it is necessary to define the set of auxiliary density operators $\{ \rho_n(t) \}$ and the vector $\bm{n}=\left(n_{1}, n_{2}, n_{k} \right)$. 
Here, $\rho_0(t)$ is the physical density matrix and the higher order $\rho_n(t)$ represent bath memory effects. 
It is customary within HEOM to consider the dynamics taking only the first $L$ terms of $\rho_n$. 

The time evolution of the system in HEOM for a system coupled to a single harmonic bath \cite{Bai2024} is given by: 
\begin{equation}
    \frac{d\rho_n(t)}{dt}= - \left(i\mathcal{L}_S+ \sum_{k=1}^{K} n_{k} \gamma_{k} \right)\rho_n -i \sum_{k=1} \left(\mathcal{C}_k^+ \rho_{n_k^+} + \mathcal{C}_k^- \rho_{n_k^-}\right), 
\end{equation}
with $\mathcal{L}_s$ the usual Liouvillian superoperator $\mathcal{L}_s \rho=[H_S, \rho]$ and $\mathcal{C}_k^{\pm}$
are the superoperators coupling the system and the bath. 

HEOM dynamics have been used previously in  a range of systems relevant to organic photovoltaic (OPV) devices including studying charge separation in OPVs \cite{Yan2019} and studying EET in the FMO complex\cite{Ishizaki2009}. 

To try and get a more realistic representation of the vibrational modes in a Y6 dimer, we have used two baths in our HEOM dynamics; a Drude-Lorentz (Debye) spectral density to represent the slow, classical modes and an under-damped (Lorentzian) spectral density  to represent the higher frequency, intramolecular modes. 

The Drude Lorentz bath has a spectral density
     \begin{equation}
        J(\omega) = \frac{2\lambda \gamma \omega}{\gamma^2 + \omega^2}, 
\end{equation}
where in this work we use a value of $\lambda=0.034 $ eV taken from \cite{giannini2024role} and $\gamma=0.05 \ \mathrm{eV} \simeq 2 k_B T $. 

While the under-damped bath has a spectral density 
\begin{equation}
    J(\omega) = \frac{2 \Delta^2 W \omega }{(\omega^2-\omega_0^2)^2+\omega^2W^2}
\end{equation}
with $\Delta =\sqrt{2\lambda_u u_{max}^2}$, $W=\gamma_u$, $\omega_0=u_{max}$. In this work, we use $\lambda=0.052$ eV taken from \cite{giannini2024role}, and choose $\gamma_u=0.015$ eV and $u_{max}=0.16$ eV in order to obtain convergent dynamics.  

Although HEOM is exact in principle, the infinite hierarchy of equations must be truncated so that they are computationally tractable. $L$ and $K$ are the integers which determine where the equations are truncated, with $L$ specifying the number of auxiliary density operators which are considered and $K$ determining how many terms of the Matsubara expansion are kept for a given correlation function. We find that, for our system, HEOM converges by $L=3$, $K=2$ if we consider only the Drude-Lorentz bath (see figure \ref{HEOM_Con} in section \ref{con_HEOM}). We therefore consistently truncate at $L=3$. For the Drude-Lorentz bath, we truncate the Matsubara expansion at $K=3$, with the slightly larger $K$ value allowing for any effects on convergence from the fast-mode bath. Due to the rapid decay of the fast mode's correlation function, we keep only the first term in its Matsubara expansion (i.e, $K=1$), as is standard practise.

In addition to the above terms in HEOM dynamics, we include a dissipater term which enables relaxation of the singlet excited states to the ground state, via e.g. fluorescence. In our dynamics, we choose the FE(S)-S0 dissipater rate to be $10^9 s^{-1}$ and the CT(S)-S0 dissipater rate to be $10^8 s^{-1}$, based on experimental data from \cite{Zou2020}.

\section{HEOM projection operator}

The choice of projection operator determines the states between which one calculates the effective rates. Thus, to match the results of our HEOM dynamics, we use a projection operator which selects the effective states of $H_\mathrm{eff}$. 
This means that the action of $\mathcal{P}$ on the set of auxiliary density operators is 

\begin{equation}
    \mathcal{P}\rho = (\sum_i   \ket{I}\braket{I|\rho_0|I}\bra{I} ,\{0 ; n \neq (0,0,0)\})^T
\end{equation}

where we have written $\{\rho_n\}$ as a column vector of the form $(\rho_0,\{\rho_n ; n \neq (0,0,0)\})^T$ and the $\ket{I}$ are the effective states defined in Section \ref{subsec:DefineHam}.

\section{Convergence of HEOM dynamics}\label{con_HEOM}

\begin{figure}
    \centering
    \includegraphics[width=1\linewidth]{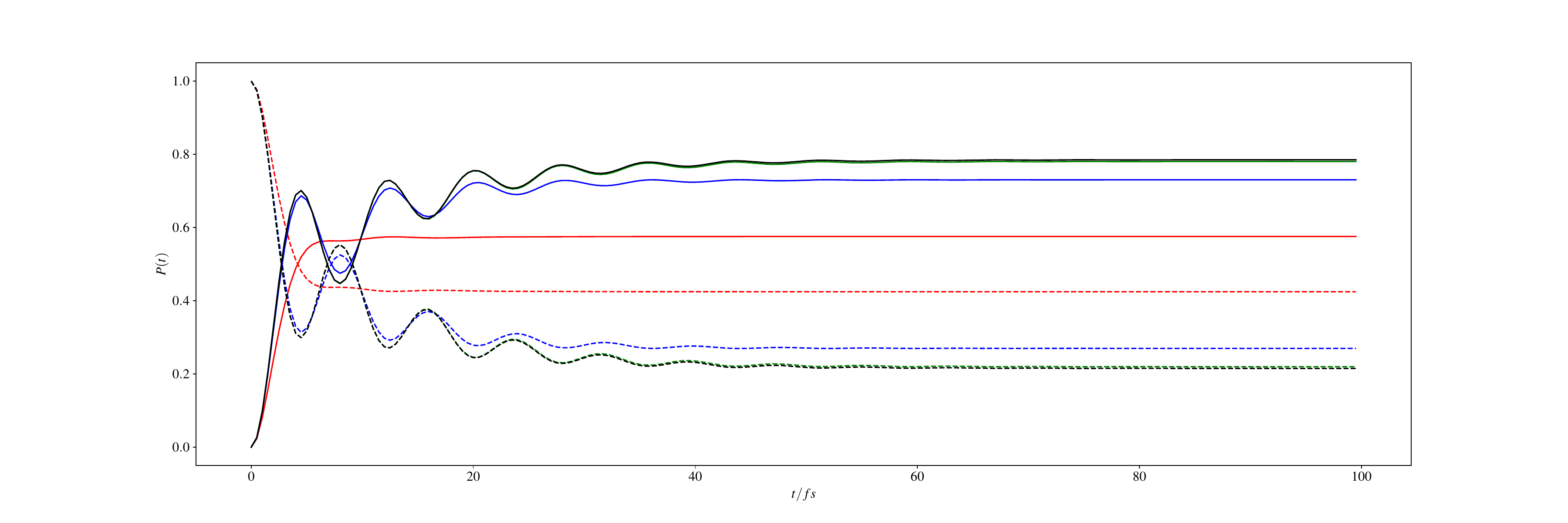}
    \caption{The convergence of HEOM dynamics with different $L$ and $K$ values. Dotted line shows FE state, solid line shows CT state. $L=3, K=0$ (red), $L=3$, $K=1$ (blue), $L=3, K=2$ (green), $L=4, K=2$ (black). Clear evidence of convergence by $L=3, K=2$.}
    \label{HEOM_Con}
\end{figure}

In order to run HEOM dynamics, we needed to check the convergence with parameters $L$ and $K$. 
Figure \ref{HEOM_Con} shows that the Y6 dimers HEOM dynamics has converged for $L=3$ and $K=2$.  This dynamics was simulated with the pyrho code\cite{pyrho}. 

In Figures \ref{HEOM_diss_Y6} and \ref{HEOM_diss_Y6Se} we show HEOM dynamics without the inclusion of the dissipater describing decay from the singlet excited states to the ground state. We find that the triplet yield obtained at 1 ns is not significantly impacted by the presence of the dissipater term in the dynamics. Physically this is reasonable as FE(S)-CT(S) transfer is very rapid (on ${\cal{O}}(100fs)$) and so the decay from FE(S) using the dissipater has limited impact on the yield of CT(S) at 100 fs. Additionally, CT(S) dissipation is slow relative to the timescale of our simulation and so has a limited impact on the final yield of FE(T).

\section{Extracting the Initial Rate of Triplet State Formation from the Effective Rates}

\begin{figure}
    \centering
    \includegraphics[width=1\linewidth]{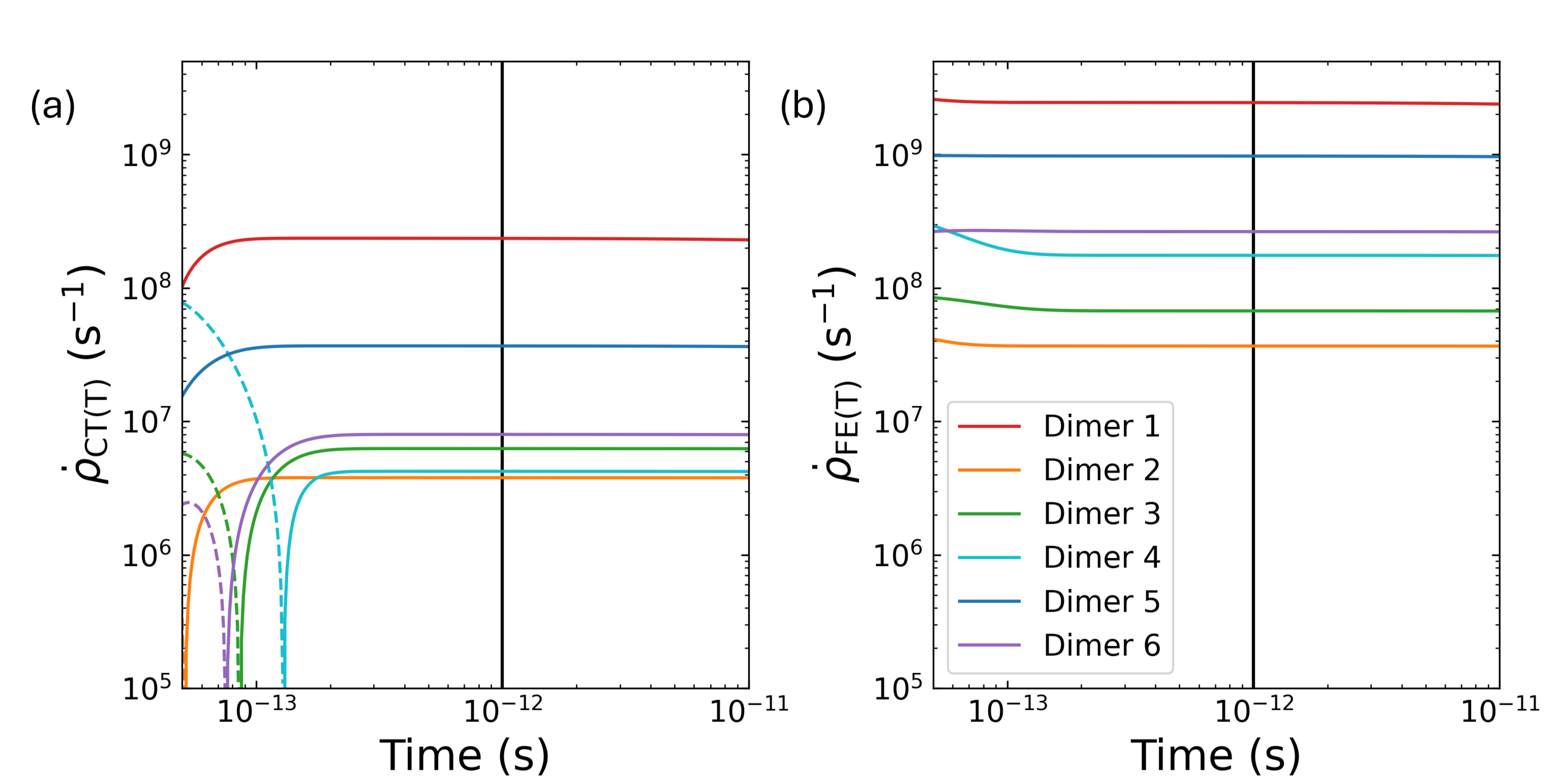}
    \caption{Net rate of change of the population density for (a) the CT(T) state and (b) the FE(T) state for all Y6Se dimers, as calculated using the HEOM-derived effective rates. The black line indicates the time point at which the data were sampled to extract the effective rates of state formation.}
    \label{TripletFormationRates_YSe}
\end{figure}

From the dynamics calculated using the HEOM-derived effective rates, we can extract the rate of CT(T) and FE(T) formation on the ps timescale.To do this, we first calculate $dP/dt$ for each population as a function of time, as shown in the Supplementary Information, Figures \ref{TripletFormationRates_Y6} and \ref{TripletFormationRates_YSe}. 
For the CT(T) and FE(T) states, $dP/dt$ has a constant value over the range \SIrange{0.3}{10}{\pico\second}, corresponding to populations which are increasing mono-exponentially. 
By sampling the value of $dP/dt$ at \SI{1}{\pico\second}, we can thus get the effective rate of FE(T) and CT(T) formation including contributions from all possible pathways. These values are listed in Table \ref{Initialrate_differenttypes}. 

\begin{figure}
    \centering
    \includegraphics[width=1\linewidth]{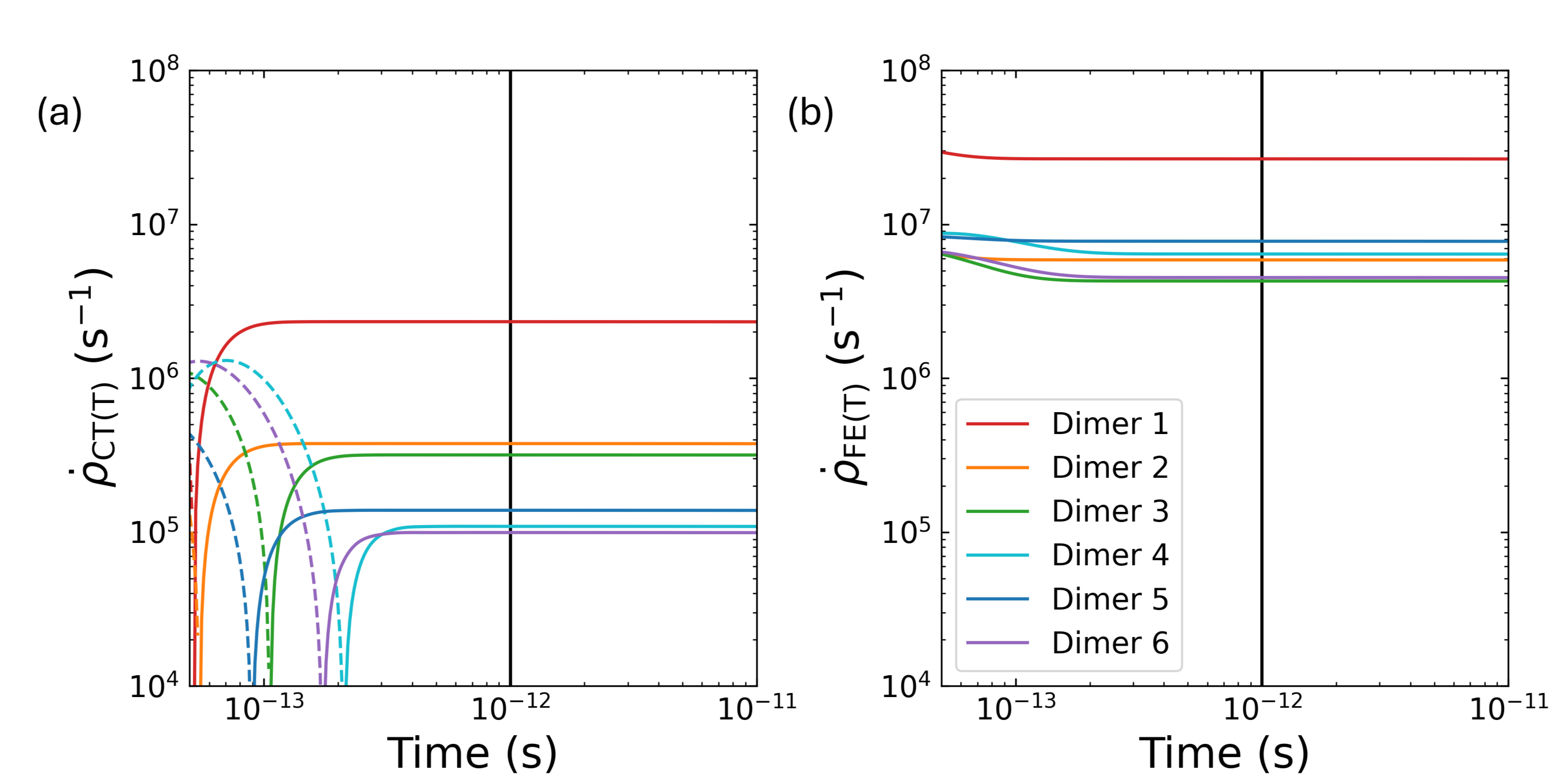}
    \caption{Net rate of change of the population density for (a) the CT(T) state and (b) the FE(T) state for all Y6 dimers, as calculated using the HEOM-derived effective rates. The black line indicates the time point at which the data were sampled to extract the effective rates of state formation.}
    \label{TripletFormationRates_Y6}
\end{figure}

\section{HEOM dynamics for other dimers}

\begin{figure}
    \centering
    \includegraphics[width=1\linewidth]{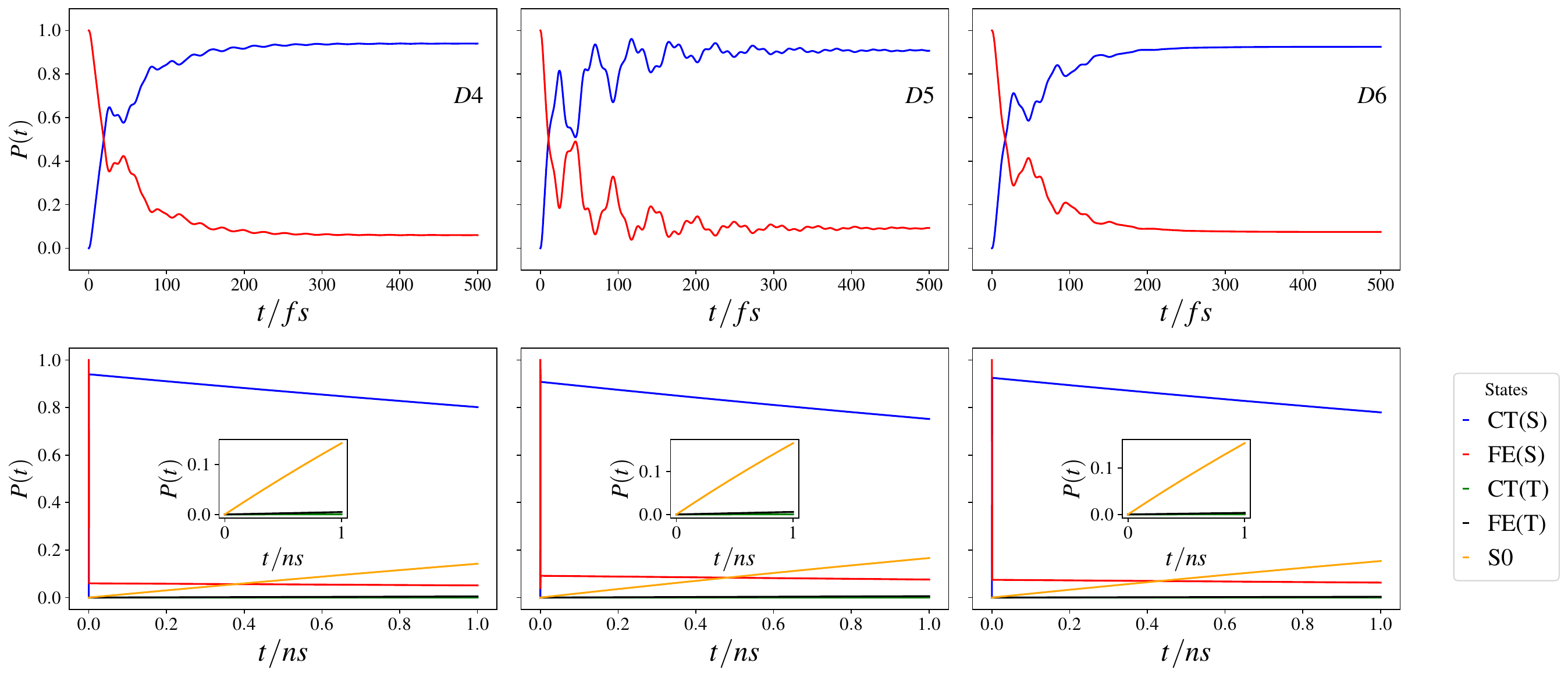}
    \caption{The results for the dynamics obtained using HEOM for the Y6 dimers (shown left to right) D4, D5 and D6 following photoexcitation into the singlet FE state at $t=0$. The top panels show the short time dynamics (${\cal{O}}(500fs)$) and bottom panels show the longer time dynamics (${\cal{O}}(1ns))$. The different types of state are as indicated in the legend. Inset shows the long-time amount of triplets formed. }
    \label{HEOM_Y6_sup}
\end{figure}

\begin{figure}
    \centering
    \includegraphics[width=\linewidth]{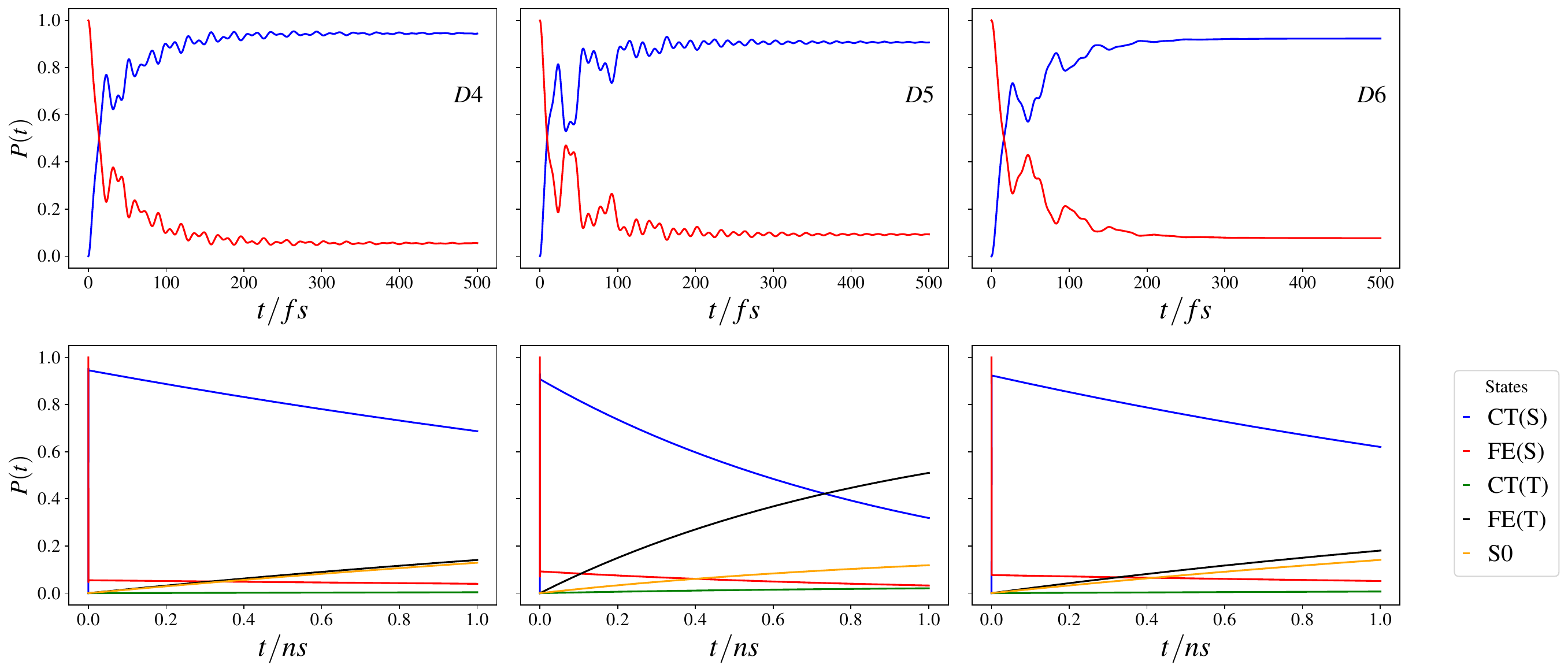}
    \caption{The results for the dynamics obtained using HEOM for the Y6Se dimers (shown left to right) D4, D5 and D6 following photoexcitation into the singlet FE state at $t=0$. The top panels show the short time dynamics (${\cal{O}}(500fs)$) and bottom panels show the longer time dynamics (${\cal{O}}(1ns))$. The different types of state are as indicated in the legend.}
    \label{HEOM_Y6Se_sup}
\end{figure}

Figures \ref{HEOM_Y6_sup} and \ref{HEOM_Y6Se_sup} show the HEOM dynamics for the other Y6 and Y6Se dimers not shown in the main paper. 

\section{FE(T) and CT(T) Fluxes for Other Dimers}

The figures in this section show the net fluxes into the FE(T) and CT(T) states, as calculated using the HEOM-derived effective rates. 

\begin{figure}
    \centering
    \includegraphics[width=1\linewidth]{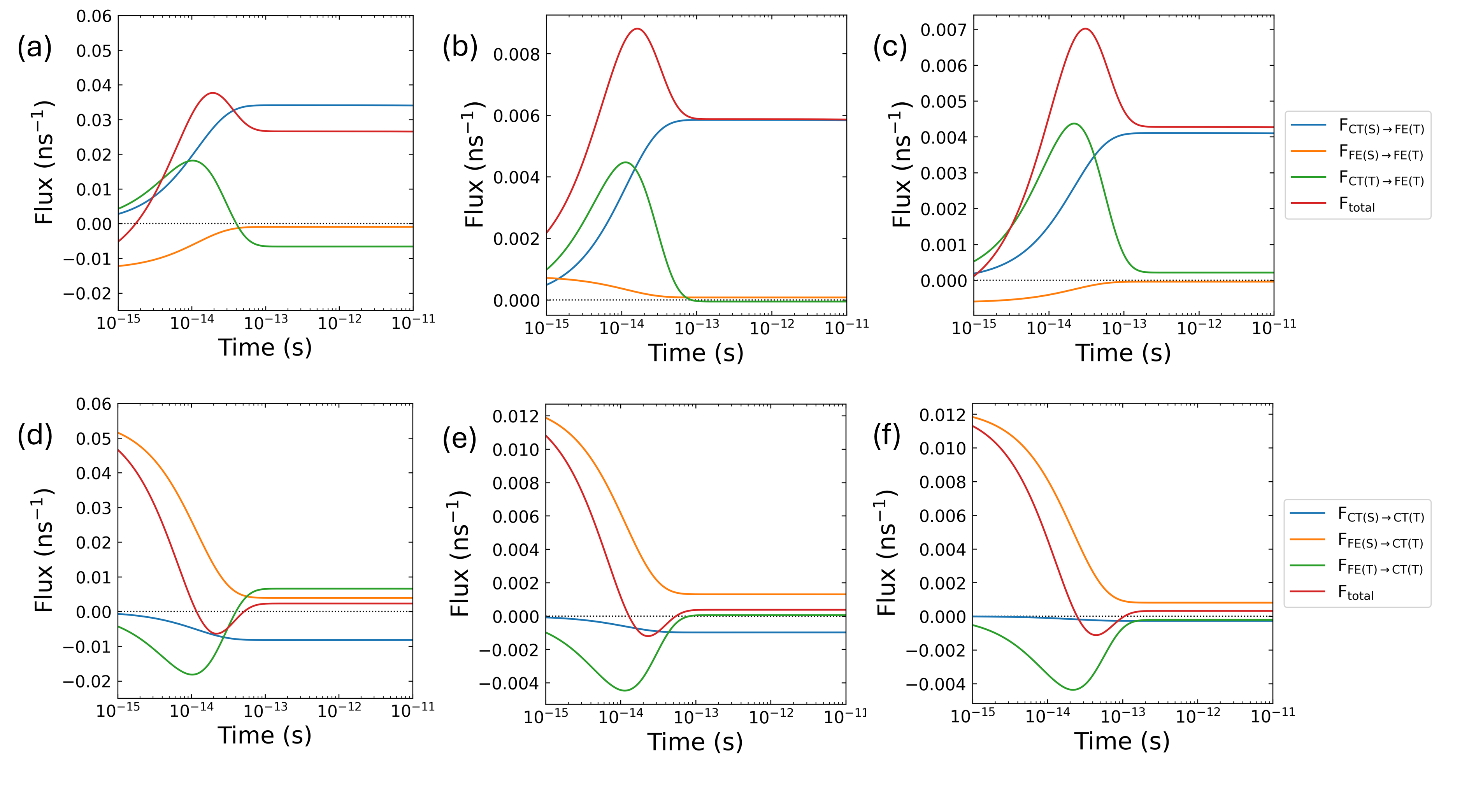}
    \caption{The net population fluxes into (a-c) the FE(T) state and (d-f) the CT(T) state for Y6 dimers D1 (a,d), D2 (b,e), and D3 (c,f). The different fluxes are as indicated in the legend, with $F_\mathrm{total}$ being the total net flux into the state.}
    \label{Fluxes_Y6_D13}
\end{figure}

\begin{figure}
    \centering
    \includegraphics[width=1\linewidth]{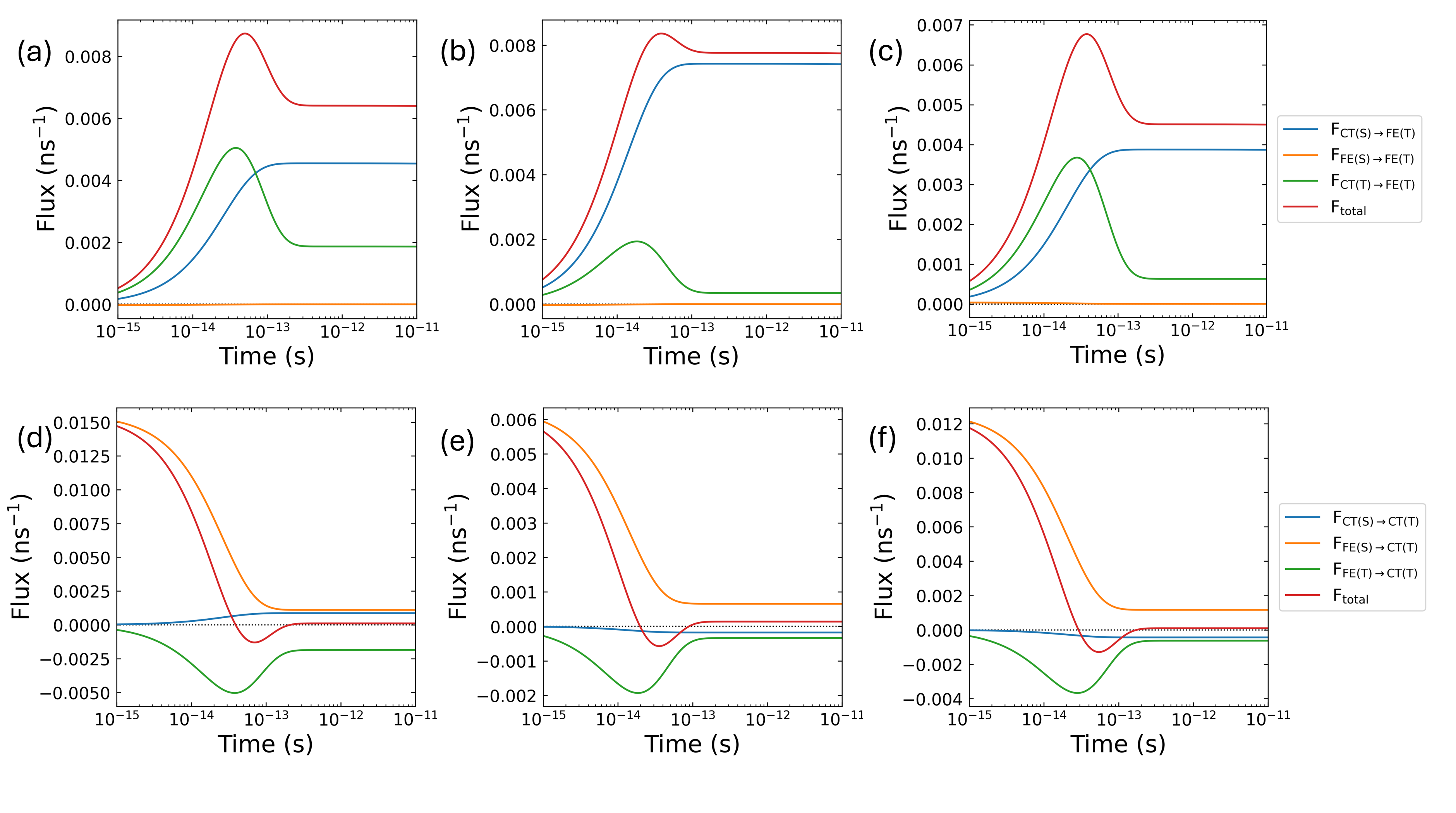}
    \caption{The net population fluxes into (a-c) the FE(T) state and (d-f) the CT(T) state for Y6 dimers D4 (a,d), D5 (b,e), and D6 (c,f). The different fluxes are as indicated in the legend, with $F_\mathrm{total}$ being the total net flux into the state.}
    \label{Fluxes_Y6_D46}
\end{figure}

\begin{figure}
    \centering
    \includegraphics[width=1\linewidth]{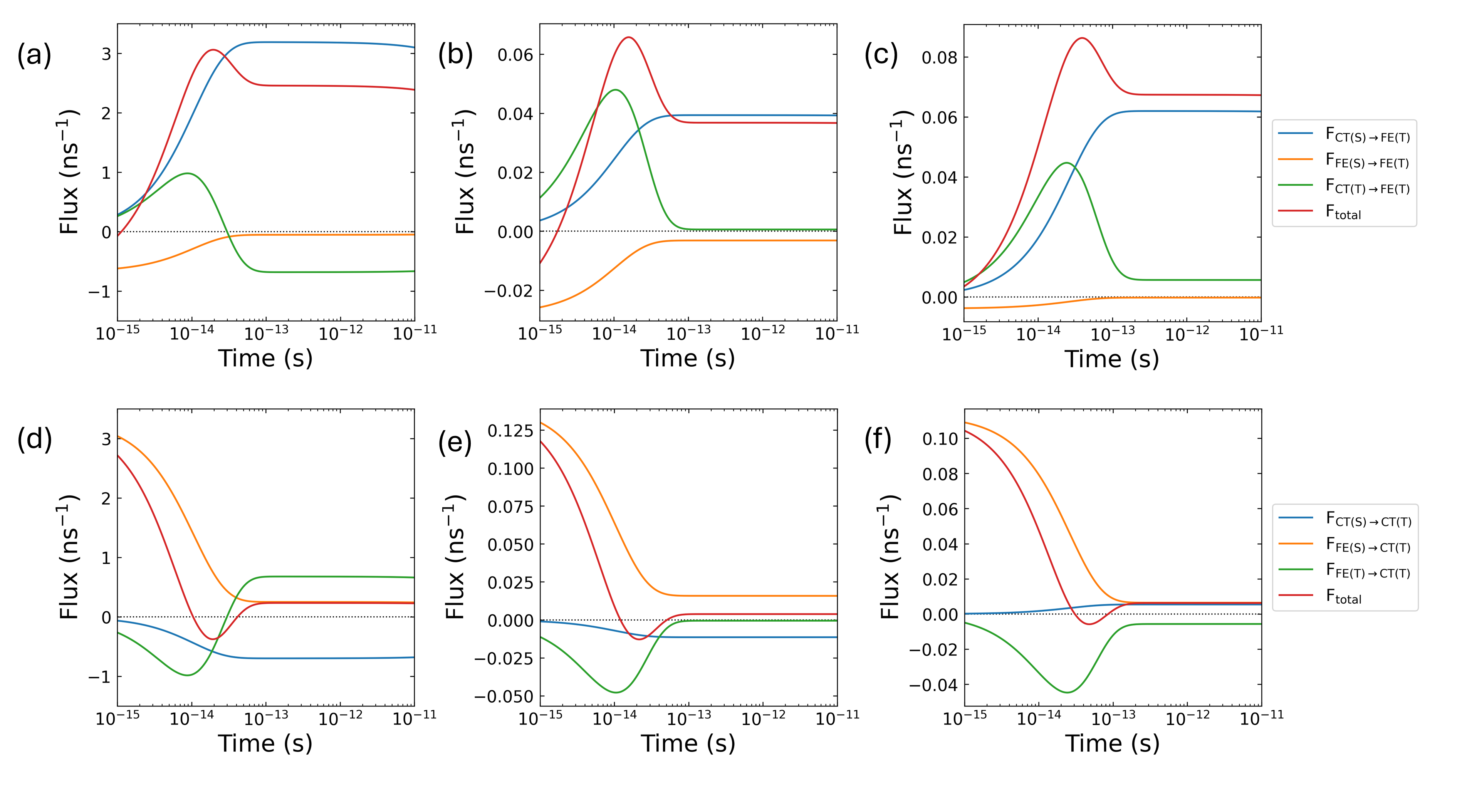}
    \caption{The net population fluxes into (a-c) the FE(T) state and (d-f) the CT(T) state for Y6Se dimers D1 (a,d), D2 (b,e), and D3 (c,f). The different fluxes are as indicated in the legend, with $F_\mathrm{total}$ being the total net flux into the state.}
    \label{Fluxes_YSe_D13}
\end{figure}

\begin{figure}
    \centering
    \includegraphics[width=1\linewidth]{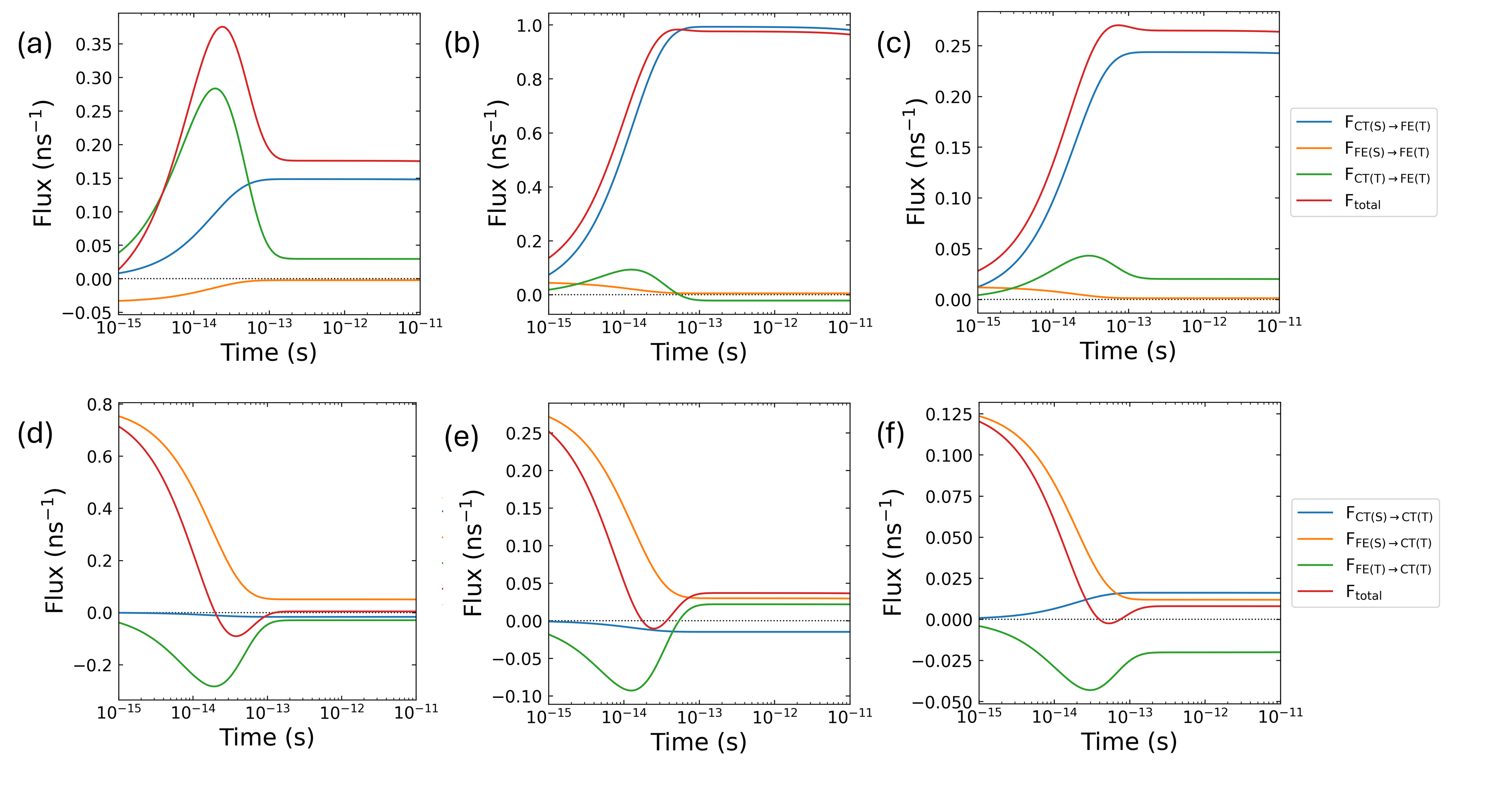}
    \caption{The net population fluxes into (a-c) the FE(T) state and (d-f) the CT(T) state for Y6Se dimers D4 (a,d), D5 (b,e), and D6 (c,f). The different fluxes are as indicated in the legend, with $F_\mathrm{total}$ being the total net flux into the state.}
    \label{Fluxes_YSe_D46}
\end{figure}

\section{HEOM dynamics Monomer \textit{vs} Dimer}

\begin{figure}
    \centering
    \includegraphics[width=1\linewidth]{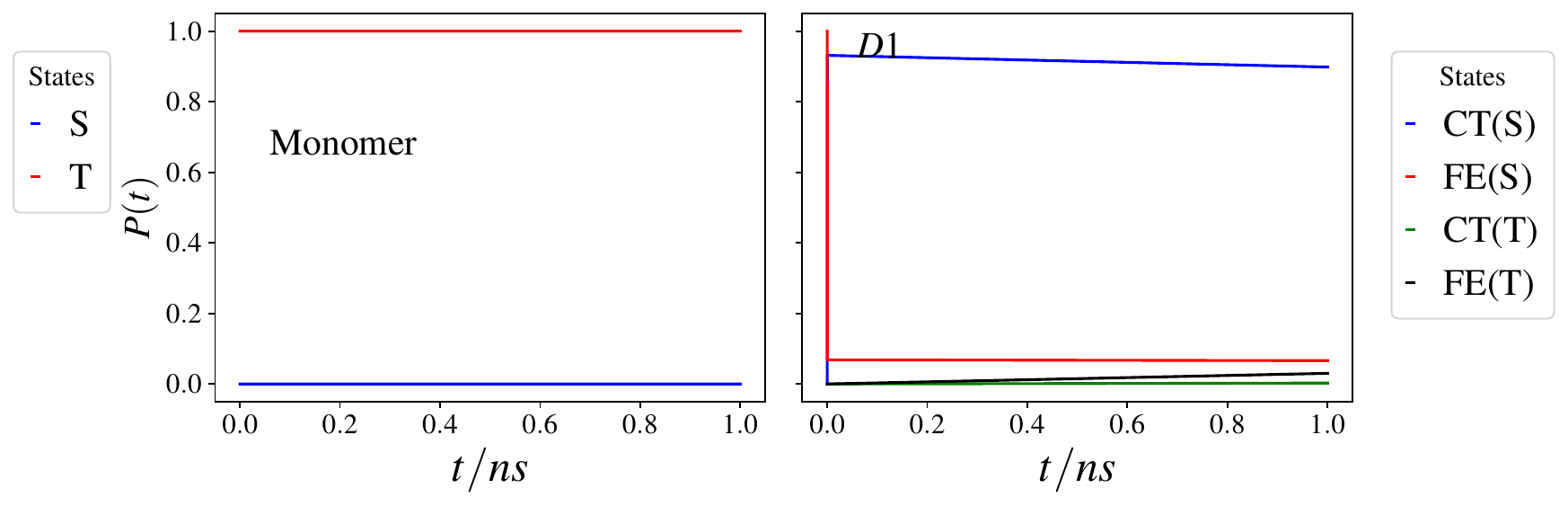}
    \caption{The results for the dynamics obtained using HEOM for D1 Y6 dimer and monomer following photoexcitation into the singlet FE state at $t=0$. The different types of state are as indicated in the legend.}
    \label{HEOM_monomer}
\end{figure}

Figure \ref{HEOM_monomer} shows the HEOM dynamics for the Y6 D1 Dimer compared to the monomer. This figure shows that negligible triplet formation is formed on a nanosecond time scale considering only the Y6 monomer. 

\section{HEOM dynamics without a dissipater}

\begin{figure}
    \centering
    \includegraphics[width=1\linewidth]{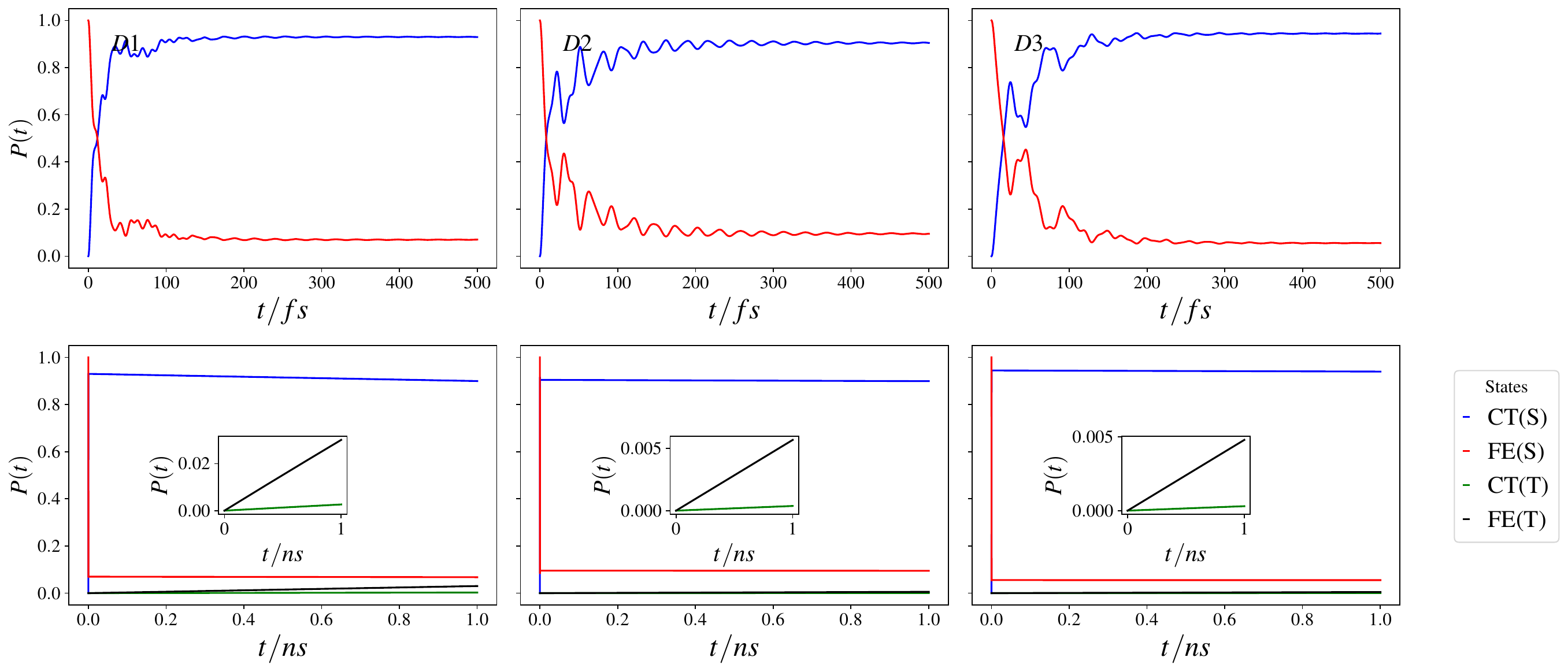}
    \caption{The results for the dynamics obtained using HEOM for the Y6 dimers without the inclusion of the dissipaters (shown left to right) D1, D2 and D3 following photoexcitation into the singlet FE state at $t=0$. The top panels show the short time dynamics (${\cal{O}}(500fs)$) and bottom panels show the longer time dynamics (${\cal{O}}(1ns))$. The different types of state are as indicated in the legend. Inset shows the long-time amount of triplets formed. }
    \label{HEOM_diss_Y6}
\end{figure}

\begin{figure}
    \centering
    \includegraphics[width=\linewidth]{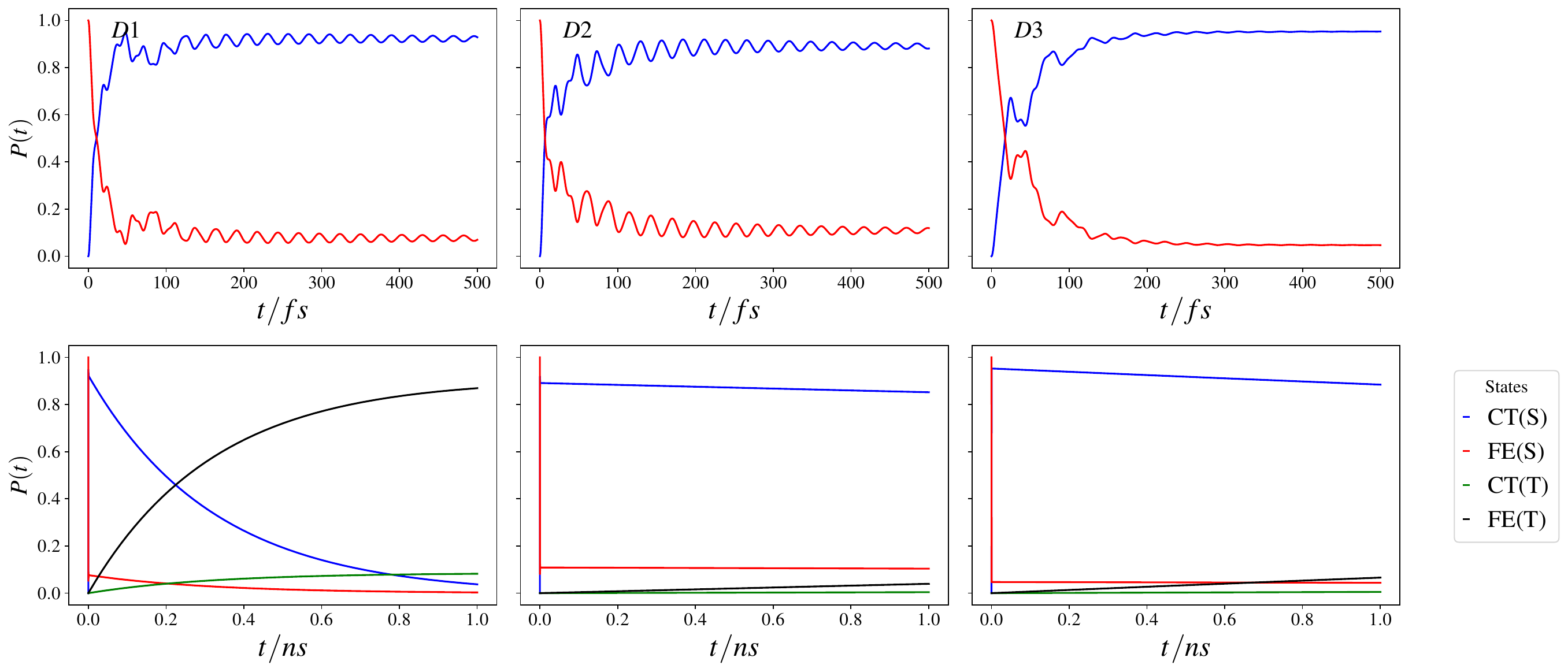}
    \caption{The results for the dynamics obtained using HEOM for the Y6Se dimers without the inclusion of the dissipaters (shown left to right) D1, D2 and D3 following photoexcitation into the singlet FE state at $t=0$. The top panels show the short time dynamics (${\cal{O}}(500fs)$) and bottom panels show the longer time dynamics (${\cal{O}}(1ns))$. The different types of state are as indicated in the legend.}
    \label{HEOM_diss_Y6Se}
\end{figure}

Figures \ref{HEOM_diss_Y6} and \ref{HEOM_diss_Y6Se} show the HEOM dynamics for the Y6 and Y6Se D1-D3 dimers without the inclusion of the dissipater term in the HEOM dynamics.

\section{Comparing HEOM dynamics to Marcus theory and effective rates}

\begin{figure}
    \centering
    \includegraphics[width=1\linewidth]{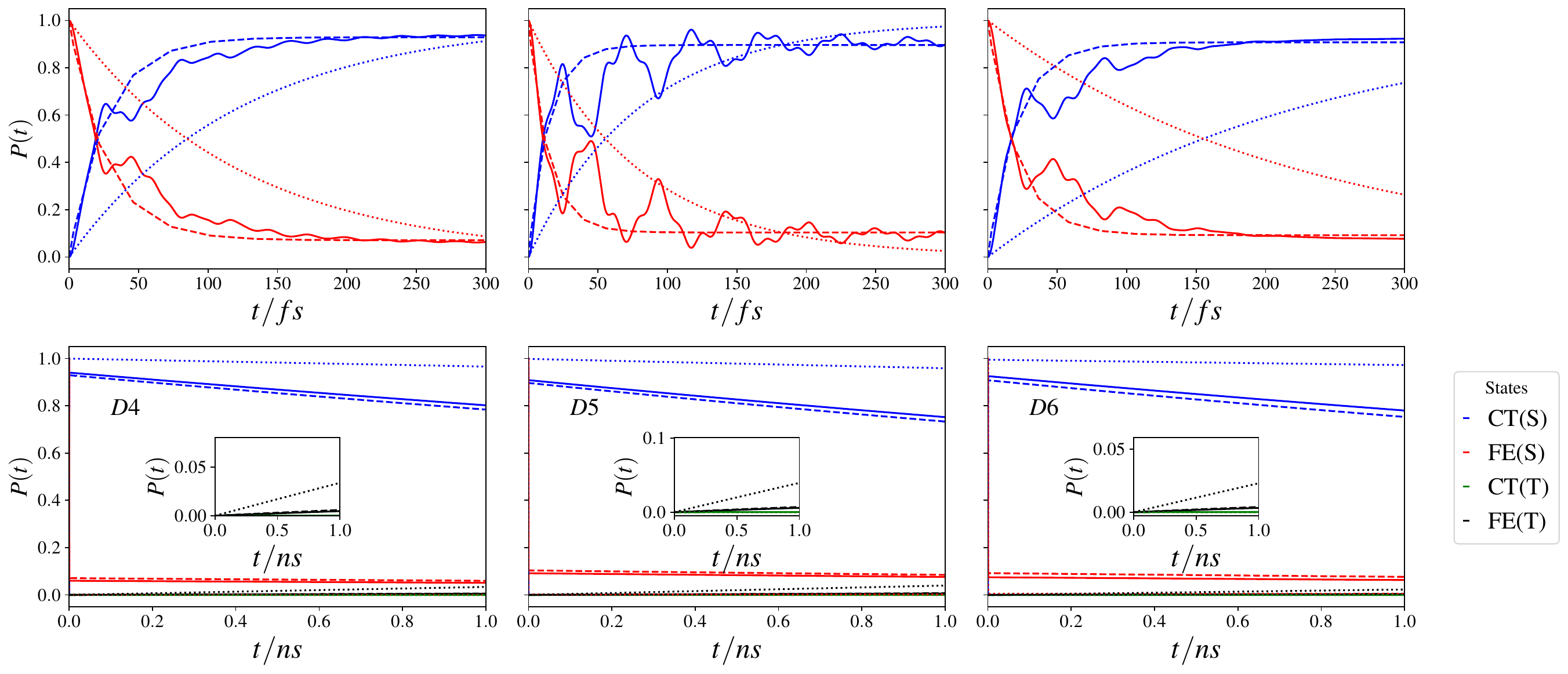}
    \caption{The results for the dynamics obtained using HEOM (bold line), effective rates from HEOM (dashed line) and Marcus theory (dotted line)  for the Y6 dimers (shown left to right) D4, D5 and D6 following photoexcitation into the singlet FE state at $t=0$. The top panels show the short time dynamics (${\cal{O}}(300fs)$) and the bottom panels show the longer time dynamics (${\cal{O}}(1ns))$. The different types of state are as indicated in the legend.}
    \label{method_compare_od}
\end{figure}

\begin{figure}
    \centering
    \includegraphics[width=1\linewidth]{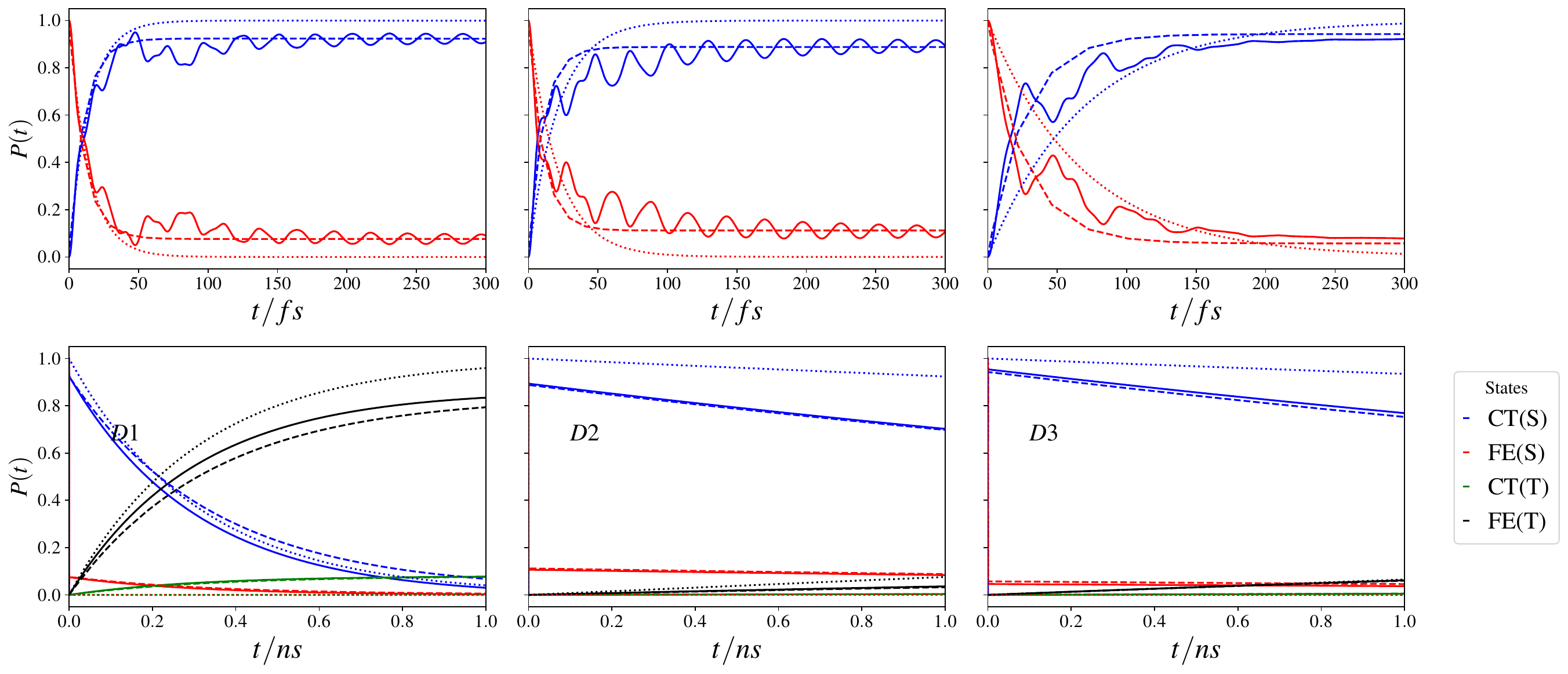}
    \caption{The results for the dynamics obtained using HEOM (bold line), effective rates from HEOM (dashed line) and Marcus theory (dotted line)  for the Y6Se dimers (shown left to right) D1, D2 and D3 following photoexcitation into the singlet FE state at $t=0$. The top panels show the short time dynamics (${\cal{O}}(300fs)$) and the bottom panels show the longer time dynamics (${\cal{O}}(1ns))$. The different types of state are as indicated in the legend.}
    \label{method_compare_Se1}
\end{figure}

\begin{figure}
    \centering
    \includegraphics[width=1\linewidth]{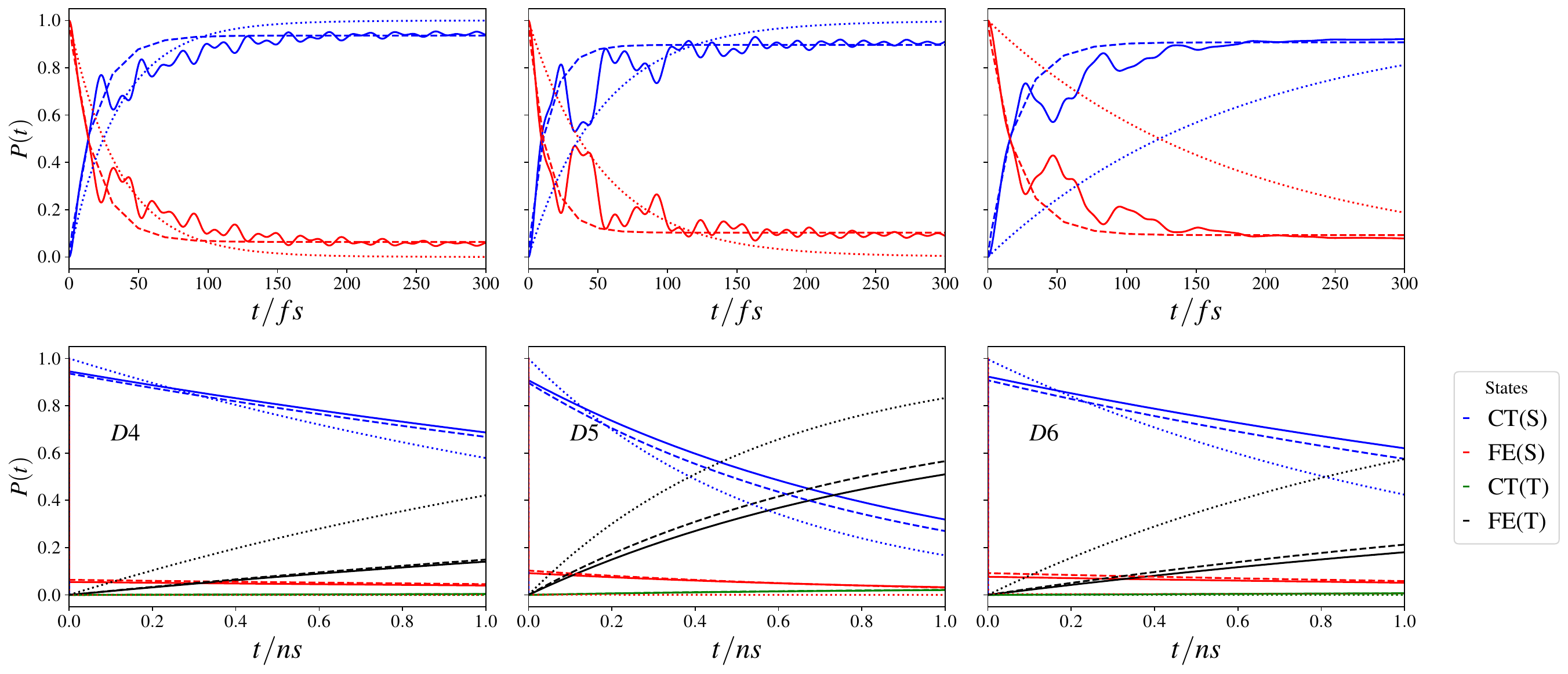}
    \caption{The results for the dynamics obtained using HEOM (bold line), effective rates from HEOM (dashed line) and Marcus theory (dotted line)  for the Y6Se dimers (shown left to right) D4, D5 and D6 following photoexcitation into the singlet FE state at $t=0$. The top panels show the short time dynamics (${\cal{O}}(300fs)$) and the bottom panels show the longer time dynamics (${\cal{O}}(1ns))$. The different types of state are as indicated in the legend.}
    \label{method_compare_Se2}
\end{figure}

Figures \ref{method_compare_od}, \ref{method_compare_Se1} and \ref{method_compare_Se2} show how the dynamics obtained using HEOM simulations compares to the dynamics obtained using effective rates extracted from HEOM and Marcus theory for other Y6 and Y6Se dimers not shown in the main body of the text.

\section{e-h analysis singlet and triplet states Y6 dimers}

\begin{figure}
    \centering
    \includegraphics[angle=270,width=1\linewidth]{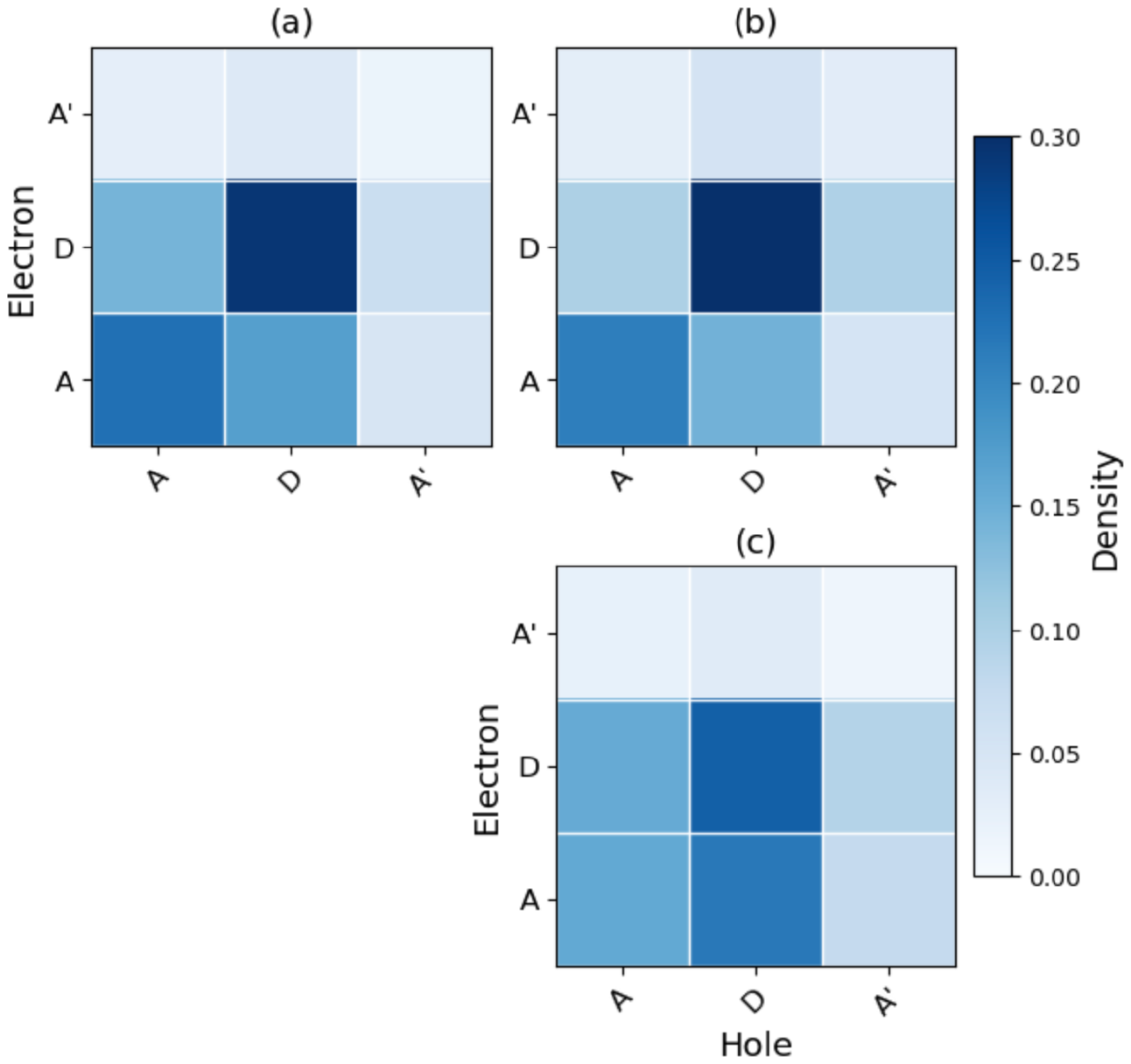}
    \caption{Electron/Hole correlation plots of the $T_1$ (a) $T_2$ (b), $S_1$ (c)  states  on the monomer. The monomer is fragmented into two (acceptor) end units and a core (donor) unit, denoted as $A/A'$ and $D$ respectively. Results obtained from electronic structure calculations using B3LYP/6-31G(d,p) and wavefunction analysis carried out using the Theodore3.2 package \cite{plasser2020theodore}. The colourbar again corresponds to the 1 particle transition density (given in terms of the electron and hole position) for the given excited state. On-diagonal elements correspond to the electron and hole being correlated on the same fragment, e.g.\ a Frenkel Exciton and off-diagonal elements correspond to Charge-Transfer excitation. }
    \label{monomer_e_h}
\end{figure}

Figure \ref{monomer_e_h} shows the electron-hole correlation plots for the $T1$ (a), $T_2$ (b) and $S_1$ (c) states of the Y6 monomer. This figure shows that the distribution of the electron and hole is very similar for all three electronic states,  with the $T_1$ and $S_1$ states being slightly more similar than the $T_2$ and $S_1$ states. From El-Sayed's rule we expect the SOCME's between the $S_1$ state and both the $T_1$ and $T_2$ states to be small.   This is exactly what we observe numerically, where SOCME between the $S_1$ and $T_1$ states is $0.014cm^{-1}$, and  between the $S_1$ and $T_2$ is $0.07cm^{-1}$.  

\begin{figure}
    \centering
    \includegraphics[, angle=270, width=1\linewidth]{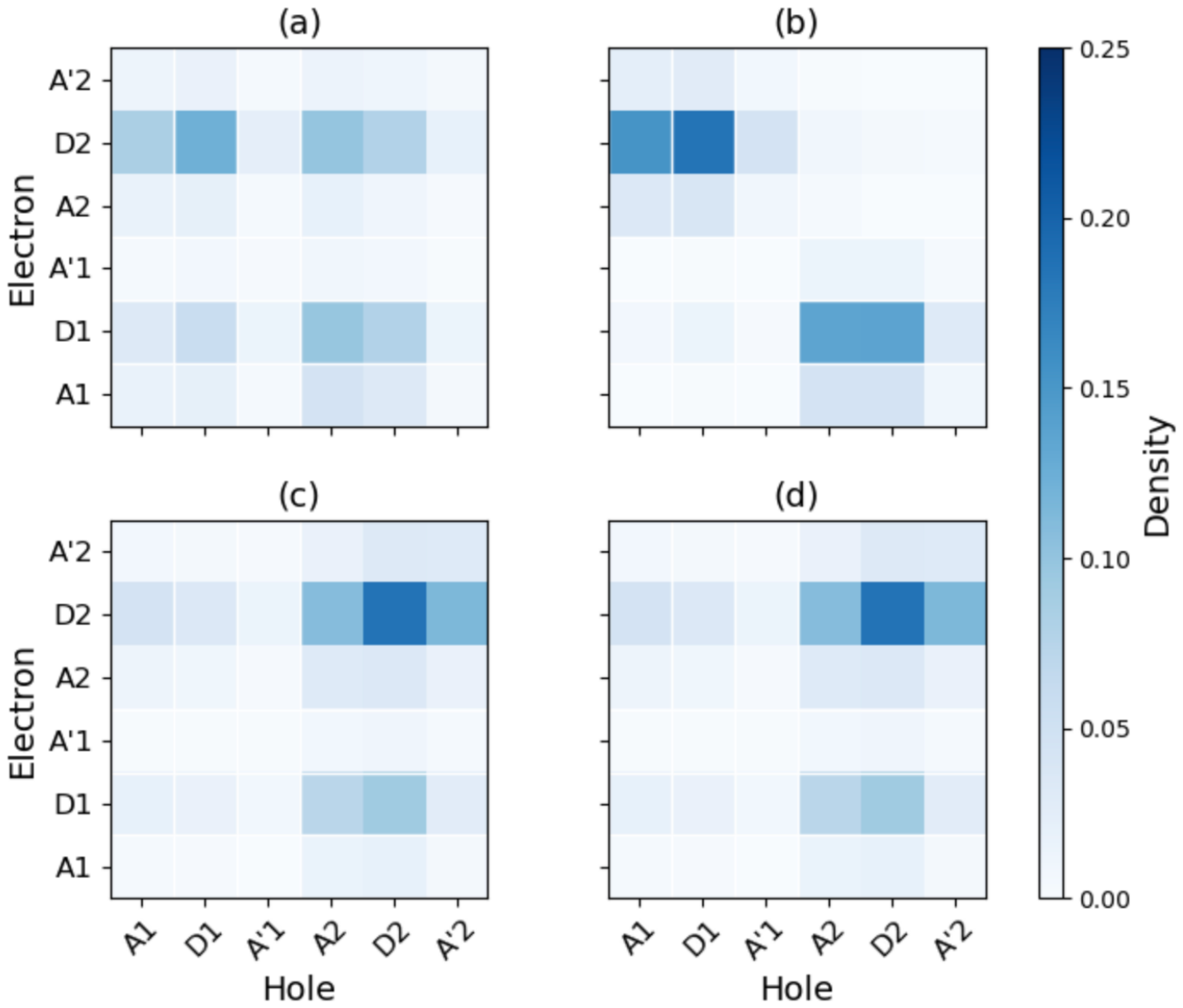}
    \caption{Electron/Hole correlation plots of the $S_1$/CT (a) $S_2$/CT (b), $S_3$/FE (c) and $S_4$/FE (d)  states  on the D1 dimer. In each dimer the monomers are fragmented into two (acceptor) end units and a core (donor) unit, denoted as $A/A'$ and $D$ respectively. Results obtained from electronic structure calculations using B3LYP/6-31G(d,p) and wavefunction analysis carried out using the Theodore3.2 package \cite{plasser2020theodore}. The colourbar again corresponds to the 1 particle transition density (given in terms of the electron and hole position) for the given excited state. On-diagonal elements correspond to the electron and hole being correlated on the same fragment, e.g.\ a Frenkel Exciton and off-diagonal elements correspond to Charge-Transfer excitation.}
    \label{D1_e_h_S}
\end{figure}

\begin{figure}
    \centering
    \includegraphics[angle=270,width=1\linewidth]{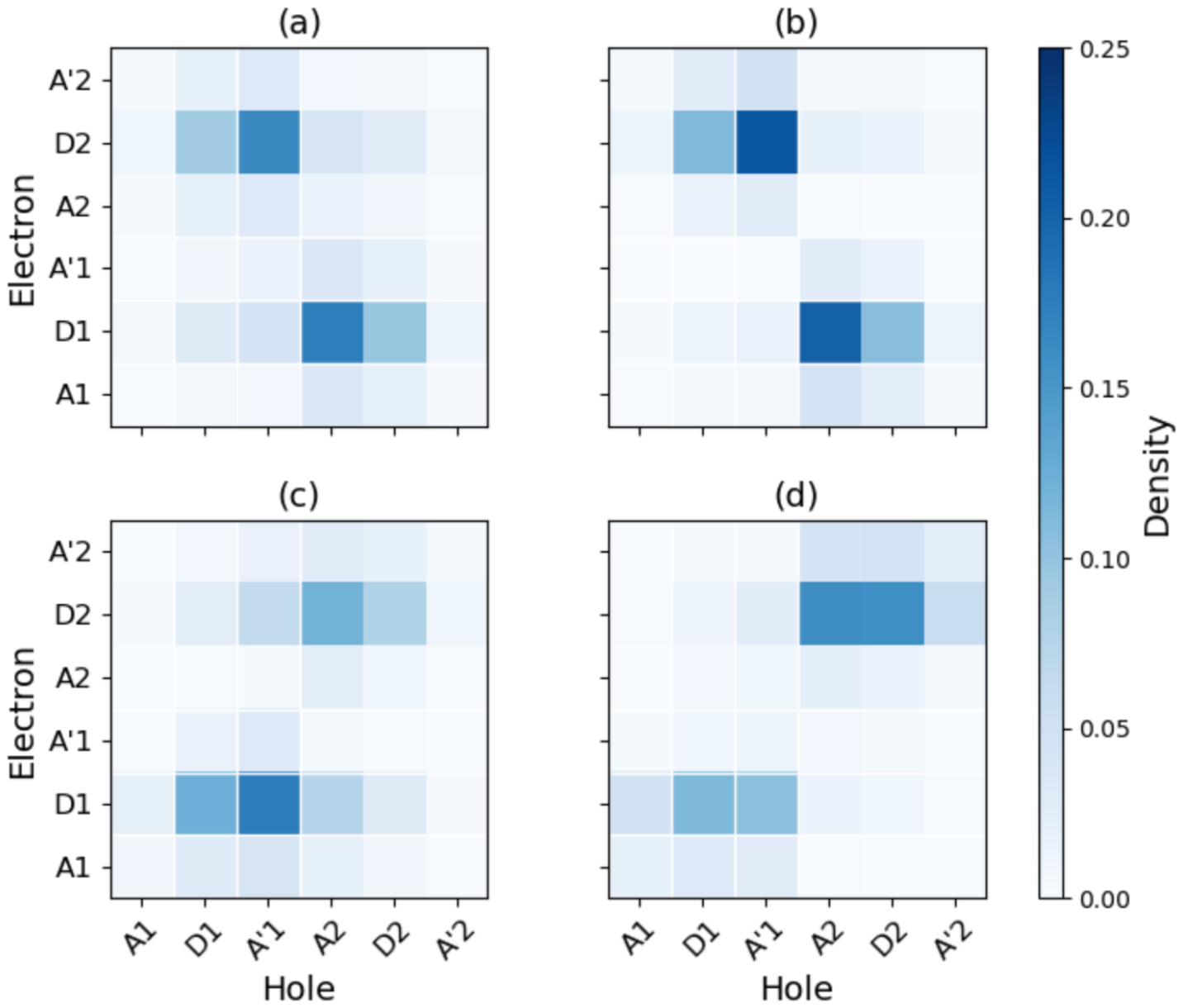}
    \caption{Electron/Hole correlation plots of the $S_1$/CT (a) $S_2$/CT (b), $S_3$/FE (c) and $S_4$/FE (d)  states  on the D2 dimer. In each dimer the monomers are fragmented into two (acceptor) end units and a core (donor) unit, denoted as $A/A'$ and $D$ respectively. Results obtained from electronic structure calculations using B3LYP/6-31G(d,p) and wavefunction analysis carried out using the Theodore3.2 package \cite{plasser2020theodore}. The colourbar again corresponds to the 1 particle transition density (given in terms of the electron and hole position) for the given excited state. On-diagonal elements correspond to the electron and hole being correlated on the same fragment, e.g.\ a Frenkel Exciton and off-diagonal elements correspond to Charge-Transfer excitation.}
    \label{D2_e_h_S}
\end{figure}

\begin{figure}
    \centering
    \includegraphics[angle=270,width=1\linewidth]{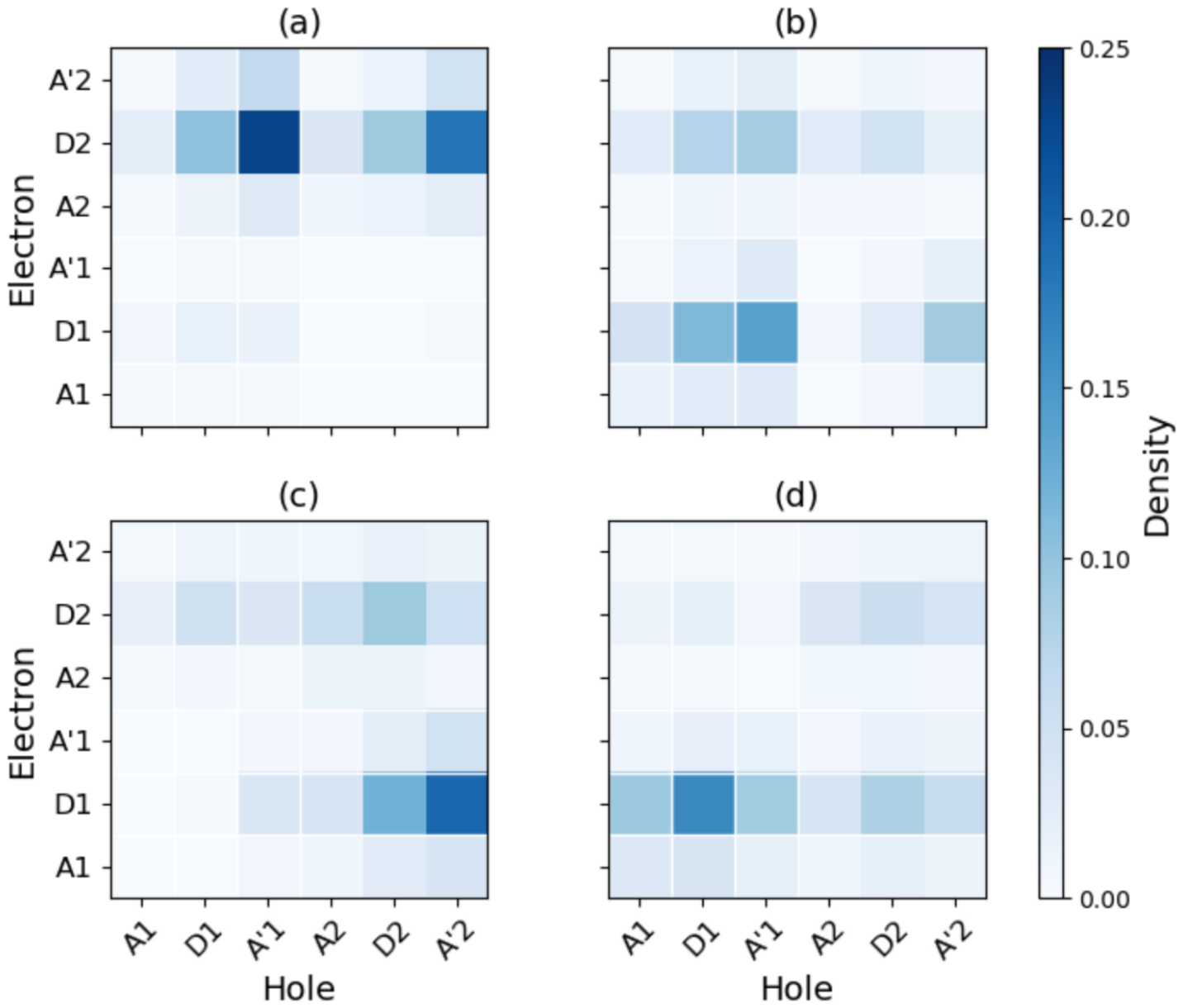}
    \caption{Electron/Hole correlation plots of the $S_1$ (a) $S_2$ (b), $S_3$ (c) and $S_4$ (d)  states  on the D4 dimer. In each dimer the monomers are fragmented into two (acceptor) end units and a core (donor) unit, denoted as $A/A'$ and $D$ respectively. Results obtained from electronic structure calculations using B3LYP/6-31G(d,p) and wavefunction analysis carried out using the Theodore3.2 package \cite{plasser2020theodore}. The colourbar again corresponds to the 1 particle transition density (given in terms of the electron and hole position) for the given excited state. On-diagonal elements correspond to the electron and hole being correlated on the same fragment, e.g.\ a Frenkel Exciton and off-diagonal elements correspond to Charge-Transfer excitation.}
    \label{D4_e_h_S}
\end{figure}

\begin{figure}
    \centering
    \includegraphics[width=1\linewidth]{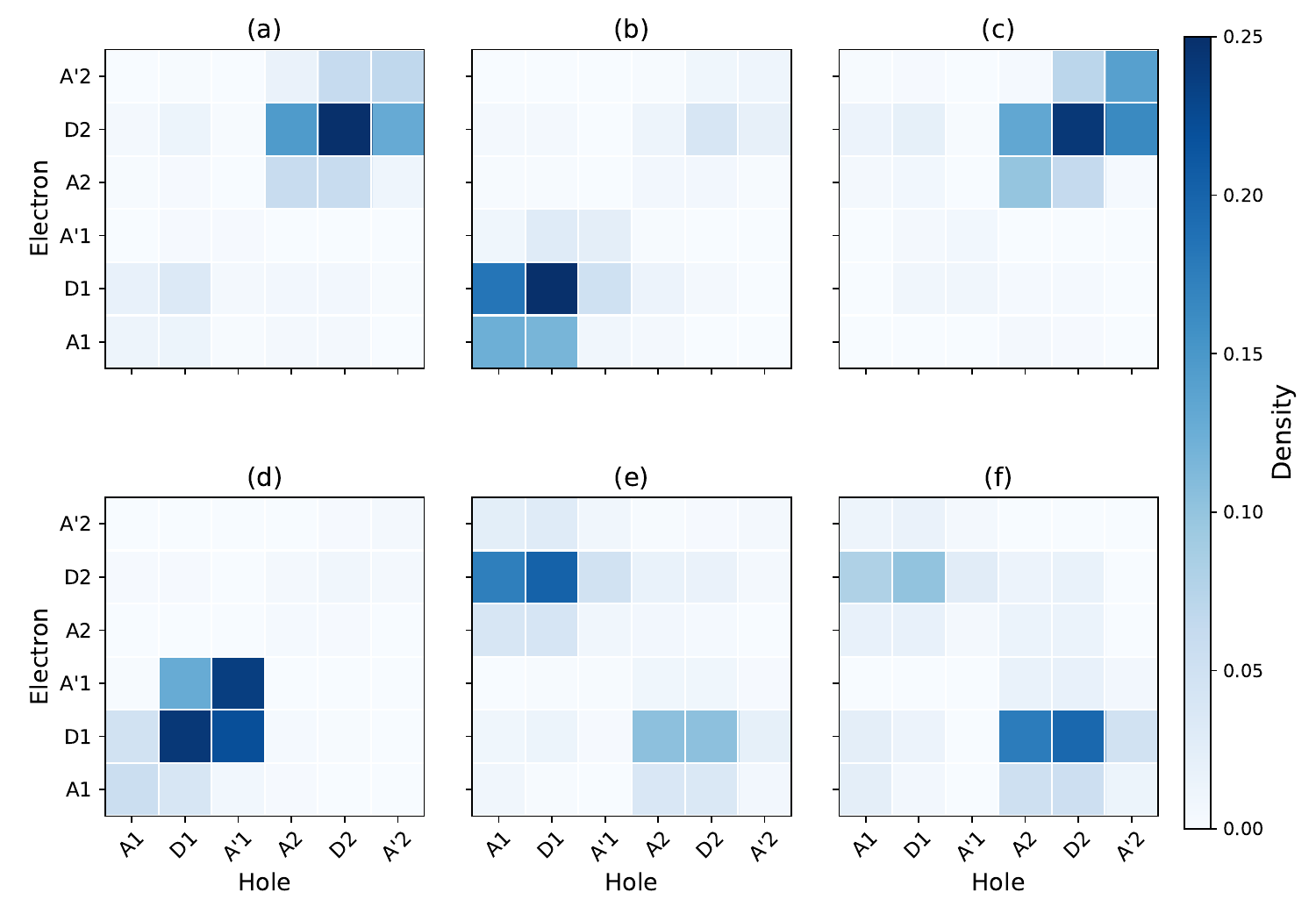}
    \caption{Electron/Hole correlation plots of the $T_1$/FE (a) $T_2$/FE (b), $T_3$/F (c), $T_4$ (d), $T_5$ (e) and $T_6$ (f) states on the D1 dimer. In each dimer the monomers are fragmented into two (acceptor) end units and a core (donor) unit, denoted as $A/A'$ and $D$ respectively. Results obtained from electronic structure calculations using B3LYP/6-31G(d,p) and wavefunction analysis carried out using the Theodore3.2 package \cite{plasser2020theodore}. The colourbar again corresponds to the 1 particle transition density (given in terms of the electron and hole position) for the given excited state. On-diagonal elements correspond to the electron and hole being correlated on the same fragment, e.g.\ a Frenkel Exciton and off-diagonal elements correspond to Charge-Transfer excitation.}
    \label{D1_e_h_T}
\end{figure}

\begin{figure}
    \centering
    \includegraphics[width=1\linewidth]{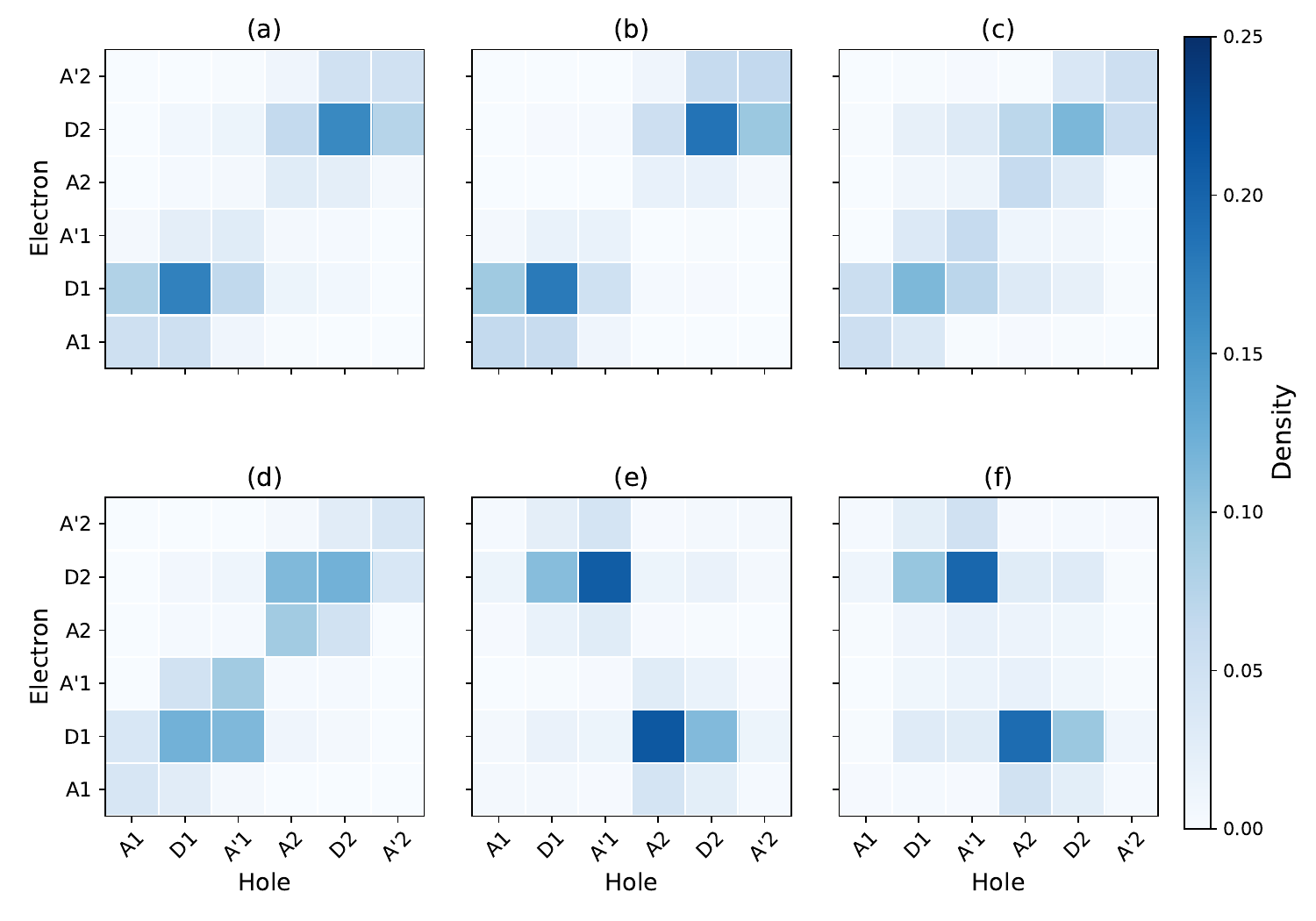}
    \caption{Electron/Hole correlation plots of the $T_1$/FE (a) $T_2$/FE (b), $T_3$/FE (c), $T_4$/FE (d), $T_5$/CT (e) and $T_6$/CT (f) states on the D2 dimer. In each dimer the monomers are fragmented into two (acceptor) end units and a core (donor) unit, denoted as $A/A'$ and $D$ respectively. Results obtained from electronic structure calculations using B3LYP/6-31G(d,p) and wavefunction analysis carried out using the Theodore3.2 package \cite{plasser2020theodore}. The colourbar again corresponds to the 1 particle transition density (given in terms of the electron and hole position) for the given excited state. On-diagonal elements correspond to the electron and hole being correlated on the same fragment, e.g.\ a Frenkel Exciton and off-diagonal elements correspond to Charge-Transfer excitation.}
    \label{D2_e_h_T}
\end{figure}

\begin{figure}
    \centering
    \includegraphics[width=1\linewidth]{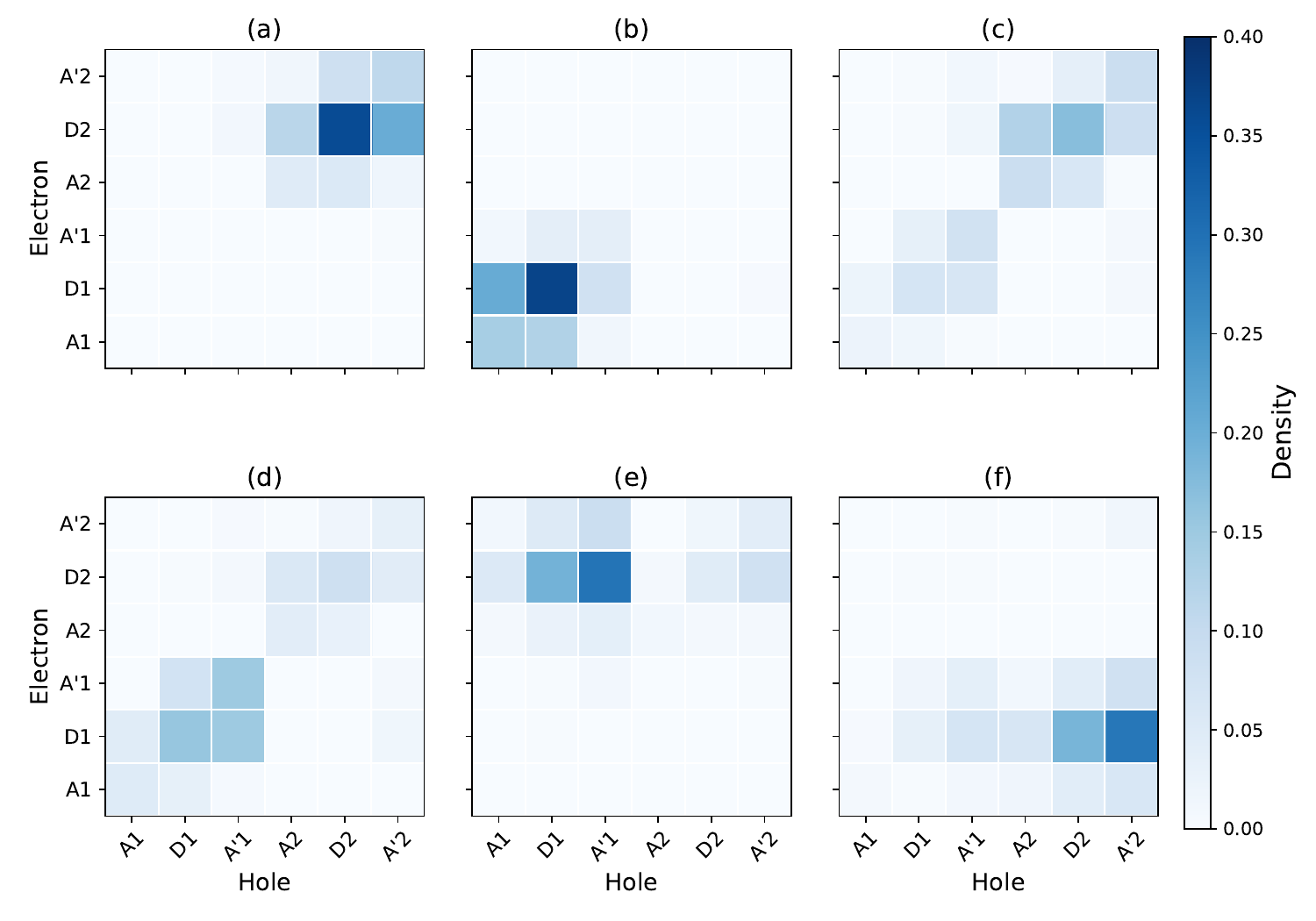}
    \caption{Electron/Hole correlation plots of the $T_1$/FE (a) $T_2$/FE (b), $T_3$/FE (c), $T_4$/FE (d), $T_5$/CT (e) and $T_6$/CT (f) states on the D4 dimer. In each dimer the monomers are fragmented into two (acceptor) end units and a core (donor) unit, denoted as $A/A'$ and $D$ respectively. Results obtained from electronic structure calculations using B3LYP/6-31G(d,p) and wavefunction analysis carried out using the Theodore3.2 package \cite{plasser2020theodore}. The colourbar again corresponds to the 1 particle transition density (given in terms of the electron and hole position) for the given excited state. On-diagonal elements correspond to the electron and hole being correlated on the same fragment, e.g.\ a Frenkel Exciton and off-diagonal elements correspond to Charge-Transfer excitation.}
    \label{D4_e_h_T}
\end{figure}

By contrast, for the dimers we see that the distribution of the electron and the hole for the singlet states (Figures \ref{D1_e_h_S}, \ref{D2_e_h_S} and \ref{D4_e_h_S} for the D1, D2 and D4 dimers) is very different to that of the the triplet states (Figures \ref{D1_e_h_T}, \ref{D2_e_h_T} and \ref{D4_e_h_T}). This reflects the different extent of localization and hybridization of the singlet and triplet states discussed in our recent paper \cite{Ward2026}. This difference in electronic distribution means all dimers SOCMES are larger than the corresponding SOCMEs in the monomer, with this difference being largest when the state type (i.e. whether the state is an FE or CT state) changes.

\end{document}